\documentclass[twocolumn,trackchanges]{aastex63} 
\usepackage{amsmath} 
\usepackage{booktabs}
\usepackage[utf8]{inputenc} 
\usepackage{xcolor} 
\usepackage{xspace}
\usepackage{graphicx} 
\usepackage{hyperref}
\usepackage[caption=false]{subfig}

\newcommand{\mymacro}[1]{\textcolor{green}{\ensuremath{#1}}}

\newcommand{\ThisEvent}{GW190521\xspace}
\newcommand{\OtherEvent}{S190521r\xspace}

\newcommand{\cWBOfflineIFAR}{\mymacro{4900} yr\xspace}

\newcommand{\cWBOfflineLHVSNR}{\mymacro{14.7}\xspace}

\newcommand{\MtotNoErrors}{\mymacro{150}}
\newcommand{\mOne}{\mymacro{85^{+21}_{-14}}}      
\newcommand{\mOneMedian}{\mymacro{85}}            
\newcommand{\mOneFivePct}{\mymacro{71}}           
\newcommand{\mTwo}{\mymacro{66^{+17}_{-18}}}   
\newcommand{\mTotal}{\mymacro{150^{+29}_{-17}}}    
\newcommand{\mChirp}{\mymacro{64^{+13}_{-8}}}    
\newcommand{\massRatio}{\mymacro{0.79^{+0.19}_{-0.29}}}    
\newcommand{\mOneDet}{\mymacro{152^{+32}_{-19}}}     
\newcommand{\mTwoDet}{\mymacro{120^{+21}_{-32}}}     
\newcommand{\mTotalDet}{\mymacro{273^{+26}_{-27}}}     
\newcommand{\mChirpDet}{\mymacro{117^{+12}_{-14}}}     

\newcommand{\mFinal}{\mymacro{142^{+28}_{-16}}}    
\newcommand{\aFinal}{\mymacro{0.72^{ +0.09}_{-0.12}}}    
\newcommand{\mTotalMinusMfinal}{\mymacro{7.6^{+2.2}_{-1.9}}}    
\newcommand{\lPeak}{\mymacro{3.7^{+0.7}_{-0.9}}}    

\newcommand{\mOneProbLessThanSixtyFive}{\mymacro{0.3}}  
\newcommand{\mOneProbLessThanFifty}{\mymacro{< 0.1}}       
\newcommand{\mTwoProbLessThanSixtyFive}{\mymacro{46.2}}     
\newcommand{\mTwoProbLessThanFifty}{\mymacro{6.6}}          
\newcommand{\mOneTwoProbInGap}{\mymacro{99.0}}              

\newcommand{\mOneProbMax}{\mymacro{12}}                   

\newcommand{\aOne}{\mymacro{0.69^{ +0.27}_{-0.62}}}       
\newcommand{\aTwo}{\mymacro{0.73^{ +0.24}_{-0.64}}}       
\newcommand{\thOne}{\mymacro{81^{+64}_{-53}}}             
\newcommand{\thTwo}{\mymacro{85^{+57}_{-55}}}             
\newcommand{\chiEff}{\mymacro{0.08^{+0.27}_{-0.36}}}      
\newcommand{\chiP}{\mymacro{0.68^{+0.25}_{-0.37}}}        
\newcommand{\logTenBFprecession}{\mymacro{+1.06\pm 0.06}} 

\newcommand{\logTenBFbbhVScusp}{\mymacro{29.8}}
\newcommand{\logTenBFbbhVSkink}{\mymacro{29.7}}

\newcommand{\cWBOoneDistance}{\mymacro{1.1}}  
\newcommand{\cWBOtwoDistance}{\mymacro{1.2}}
\newcommand{\cWBOthreeaDistance}{\mymacro{1.7}}
\newcommand{\VT}{\mymacro{9.1}}

\newcommand{\twentyFiveEventsProbGtMassOneMedian}{\mymacro{0.6\%}}
\newcommand{\twentyFiveEventsProbGtMassOneFivepct}{\mymacro{3.0\%}}
\newcommand{\fiftyEventsProbGtMassOneMedian}{\mymacro{1.3\%}}
\newcommand{\fiftyEventsProbGtMassOneFivepct}{\mymacro{5.3\%}}

\newcommand{\mOneNoHOM}{\mymacro{137}}      
\newcommand{\mOneHOM}{\mymacro{103}}        
\newcommand{\logTenBFHOM}{\mymacro{-0.38}}  

\newcommand{\skyareaBayestar}{\mymacro{1163}}
\newcommand{\skyareaLalinitial}{\mymacro{765}}
\newcommand{\skyareaLalNRSur}{\mymacro{774}}      
\newcommand{\skyareaLalIMRPhenom}{\mymacro{862}}  
\newcommand{\skyareaRIFTSEOBNR}{\mymacro{1069}}   

\newcommand{\seglen}{\mymacro{8}}
\newcommand{\flow}{\mymacro{11}}
\newcommand{\fhigh}{\mymacro{512}}
\newcommand{\samplerate}{\mymacro{1024}}

\newcommand{\ComponentMassPrior}{\mymacro{[30, 200]\,{M}_\odot}}
\newcommand{\LuminosityDistancePrior}{\mymacro{10}}

\newcommand{\LuminosityDistance}{\mymacro{5.3^{+2.4}_{-2.6}}}    
\newcommand{\LuminosityDistanceSEOBNR}{\mymacro{4.0^{+2.0}_{-1.8}}}   
\newcommand{\redshift}{\mymacro{0.82^{+0.28}_{-0.34}}}   
\newcommand{\approxredshift}{\mymacro{0.8}}  

\newcommand{\PeaktimeH}{\mymacro{1242442967.4306^{+0.0067}_{-0.0106}}}
\newcommand{\PeaktimeFiveM}{\mymacro{6.4}}
\newcommand{\PeaktimeTenM}{\mymacro{12.7}}
\newcommand{\PeaktimeFifteenM}{\mymacro{19.1}}
\newcommand{\FinalMassNRSurMedian}{\mymacro{258.3}}
\newcommand{\FreqRing}{\mymacro{66.0^{+4.0}_{-3.0}}}
\newcommand{\TauRing}{\mymacro{19.0^{+9.0}_{-7.0}}}
\newcommand{\FinalMassRing}{\mymacro{252.0^{+63.0}_{-64.0}}}
\newcommand{\FinalSpinRing}{\mymacro{0.65^{+0.22}_{-0.48}}}
\newcommand{\FinalMassTwoOvertones}{\mymacro{276.0^{+51.0}_{-50.0}}}
\newcommand{\FinalSpinTwoOvertones}{\mymacro{0.74^{+0.15}_{-0.3}}}
\newcommand{\FinalMassHM}{\mymacro{296.0^{+52.0}_{-58.0}}}
\newcommand{\FinalSpinHM}{\mymacro{0.79^{+0.12}_{-0.3}}}

\newcommand{\BayesWaveOffsourceTime}{\mymacro{4096}}
\newcommand{\BayesWaveNumOffsource}{\mymacro{200}}
\newcommand{\BayesWaveResidualSnrNinety}{\mymacro{5.97}}
\newcommand{\BayesWaveResidualPvalue}{\mymacro{0.4}}

\newcommand{\lensingMp}{\mymacro{43^{+6}_{-16}}}
\newcommand{\lensingMs}{\mymacro{35^{+11}_{-15}}}
\newcommand{\lensingz}{\mymacro{2.5^{+2.1}_{-0.7}}}
\newcommand{\lensingtau}{\mymacro{9^{+10}_{-3} \times 10^{-4}}}
\newcommand{\lensingmagnification}{\mymacro{13^{+54}_{-8}}}

\newcommand{\lensingMpb}{\mymacro{55^{+9}_{-22}}}
\newcommand{\lensingMsb}{\mymacro{44^{+14}_{-19}}}
\newcommand{\lensingzb}{\mymacro{1.8^{+1.7}_{-0.6}}}
\newcommand{\lensingtaub}{\mymacro{5^{+9}_{-3} \times 10^{-4}}}
\newcommand{\lensingmagnificationb}{\mymacro{6^{+28}_{-4}}}

\def\RelRateHalfMedian{\mymacro{-2.0}}
\def\RelRateHalfPlus{\mymacro{1.0}}
\def\RelRateHalfMinus{\mymacro{1.2}}
\def\RelRateTwoMedian{\mymacro{-4.4}}
\def\RelRateTwoPlus{\mymacro{2.0}}
\def\RelRateTwoMinus{\mymacro{2.4}}
\def\RelRateNoZeroHalfMedian{\mymacro{-2.6}}
\def\RelRateNoZeroHalfPlus{\mymacro{0.9}}
\def\RelRateNoZeroHalfMinus{\mymacro{1.1}}
\def\RelRateNoZeroTwoMedian{\mymacro{-5.5}}
\def\RelRateNoZeroTwoPlus{\mymacro{1.8}}
\def\RelRateNoZeroTwoMinus{\mymacro{2.2}}
\def\LambdaChiLow{\mymacro{0.01}}
\def\LambdaChiHigh{\mymacro{0.47}}
\def\ORNRSurNoZeroHalf{\mymacro{0.28}}
\def\ORNRSurNoZeroTwo{\mymacro{0.07}}
\def\ORIMRPHMNoZeroHalf{\mymacro{0.28}}
\def\ORIMRPHMNoZeroTwo{\mymacro{0.03}}
\def\ORNRSurHalf{\mymacro{0.86}}
\def\ORNRSurTwo{\mymacro{0.68}}
\def\ORIMRPHMHalf{\mymacro{0.81}}
\def\ORIMRPHMTwo{\mymacro{0.23}}
\def\RateMedian{\mymacro{0.13}}
\def\RatePlus{\mymacro{0.30}}
\def\RateMinus{\mymacro{0.11}}

\renewcommand{\mymacro}[1]{{\ensuremath{#1}}}

\renewcommand{\ThisEvent}{GW190521\xspace}
\newcommand{\Msun}{\ensuremath{\mathrm{M}_\odot}}

\newcommand{\editNew}[1]{{{#1}}}
\turnoffeditone

\begin{document}

\title{Properties and astrophysical implications of the \MtotNoErrors\,\Msun\ binary black hole merger \ThisEvent} 

%
\AuthorCollaborationLimit=3000  


%
\author{R.~Abbott}
\affiliation{LIGO, California Institute of Technology, Pasadena, CA 91125, USA}
\author{T.~D.~Abbott}
\affiliation{Louisiana State University, Baton Rouge, LA 70803, USA}
\author{S.~Abraham}
\affiliation{Inter-University Centre for Astronomy and Astrophysics, Pune 411007, India}
\author{F.~Acernese}
\affiliation{Dipartimento di Farmacia, Universit\`a di Salerno, I-84084 Fisciano, Salerno, Italy}
\affiliation{INFN, Sezione di Napoli, Complesso Universitario di Monte S.Angelo, I-80126 Napoli, Italy}
\author{K.~Ackley}
\affiliation{OzGrav, School of Physics \& Astronomy, Monash University, Clayton 3800, Victoria, Australia}
\author{C.~Adams}
\affiliation{LIGO Livingston Observatory, Livingston, LA 70754, USA}
\author{R.~X.~Adhikari}
\affiliation{LIGO, California Institute of Technology, Pasadena, CA 91125, USA}
\author{V.~B.~Adya}
\affiliation{OzGrav, Australian National University, Canberra, Australian Capital Territory 0200, Australia}
\author{C.~Affeldt}
\affiliation{Max Planck Institute for Gravitational Physics (Albert Einstein Institute), D-30167 Hannover, Germany}
\affiliation{Leibniz Universit\"at Hannover, D-30167 Hannover, Germany}
\author{M.~Agathos}
\affiliation{Theoretisch-Physikalisches Institut, Friedrich-Schiller-Universit\"at Jena, D-07743 Jena, Germany}
\affiliation{University of Cambridge, Cambridge CB2 1TN, UK}
% ...





\collaboration{0}{LIGO Scientific Collaboration and Virgo Collaboration}

\begin{abstract}
The gravitational-wave signal \ThisEvent 
is consistent with a binary black hole merger source at redshift \approxredshift\ 
with unusually high component masses, 
$\mOne$\,\Msun\ and $\mTwo$\,\Msun, 
compared to previously reported events, 
and shows mild evidence for spin-induced orbital precession. 
The primary falls in 
the mass gap predicted by (pulsational) pair-instability supernova theory, \editNew{in the approximate range $65 - 120\,\Msun$.
The probability that at least one of the black holes in \ThisEvent is in that range is \editNew{\mOneTwoProbInGap\%}.}
The final mass of the merger ($\mFinal\, \Msun$) classifies it as an 
intermediate-mass black hole. 
Under the assumption of a quasi-circular binary black hole coalescence, we detail the 
physical properties of \ThisEvent's source binary and its post-merger remnant, 
including component masses and spin vectors. Three different waveform models, as well as 
direct comparison to numerical solutions of general relativity, yield consistent 
estimates of these properties. 
Tests of strong-field general relativity targeting the merger-ringdown stages of 
the coalescence indicate consistency of the observed signal with theoretical predictions. 
We estimate the merger rate of similar systems to be 
$\RateMedian^{+\RatePlus}_{-\RateMinus}\,\mathrm{Gpc}^{-3}\,{}\mathrm{yr}^{-1}$.
We discuss the astrophysical implications of \ThisEvent for stellar collapse, and for the 
possible formation of black holes in the pair-instability mass gap through various channels: 
via (multiple) stellar coalescences, or via hierarchical mergers of lower-mass black holes in 
star clusters or in active galactic nuclei. 
We find it to be unlikely that \ThisEvent is a strongly lensed signal of a lower-mass 
black hole binary merger. 
We also discuss more exotic possible sources for \ThisEvent, including a highly 
eccentric black hole binary, or a primordial black hole binary. 
\end{abstract}

\keywords{Gravitational waves -- Black holes -- Intermediate-mass black holes -- Massive stars -- Supernovae}

\section{Introduction}

The gravitational-wave (GW) signal \ThisEvent \citep{GW190521-Discovery,GCN24621} was observed by the Advanced LIGO \citep{LIGOdetector} and Advanced Virgo \citep{VIRGOdetector} detectors during their third observing run (O3). 
 \editNew{The event was found with four different search pipelines, both at low-latency and offline with improved background estimation; an offline search sensitive to generic transients found \ThisEvent with a three-detector network signal-to-noise ratio of \cWBOfflineLHVSNR, and an estimated false-alarm rate of 1 in \cWBOfflineIFAR \citep{GW190521-Discovery}.}
Another candidate GW signal was reported later on the same day \citep{GCN24632}. The source of \ThisEvent is consistent with being 
a high mass binary black hole (BBH) system. 
The final merger product of \ThisEvent, with an estimated mass of {$\mFinal\,\Msun$} (all values quoted as medians with symmetric 90\% credible interval), is the first strong observational evidence for an intermediate-mass black hole (IMBH) in the mass range $10^2-10^3\,\Msun$, under the assumption of a quasi-circular BBH coalescence. 
This merger of two high mass black holes (primary mass {$\mOne\,\Msun$}, secondary mass {$\mTwo\,\Msun$}) is also exceptional as the first observation of a black hole (BH) that lies with high confidence in the mass gap predicted by pair-instability (PI) supernova theory \citep{2017ApJ...836..244W};
the probability that the primary mass is below 
$65\,\Msun$ is {$\mOneProbLessThanSixtyFive$\%}. This high component mass represents a challenge for current astrophysical formation scenarios.

The very short duration (approximately 0.1\,s) and bandwidth (around 4 cycles in the frequency band \mbox{30--80\,Hz}) of \ThisEvent means that the interpretation of the source as being a quasi-circular compact binary coalescence consisting of inspiral, merger and ringdown phases is not certain. 
Under that interpretation we find that the observed signal, including its frequency evolution, 
is entirely consistent both with three different waveform models derived from analytical and/or numerical solutions of general relativity (GR), and with direct comparisons to numerical relativity (NR) solutions.
Therefore, most of the discussion in this paper and in \citet{GW190521-Discovery} proceeds under that assumption, including our inferences on the inferred masses and spins, and on the effect of including higher-order multipoles in the waveform models. 
However, as discussed below, other interpretations are possible, adding to the exceptional nature of this event.  

Searches for IMBH binaries with total mass ${>}100\,\Msun$ and primary mass ${\lesssim} 500\,\Msun$ were carried out in data from Initial LIGO and Virgo \citep{Virgo:2012aa,aasi:2014iwa} and from the first and second observing runs of the Advanced detector era,
O1 and O2 \citep{Abbott:2017iws,2019PhRvD.100f4064A}. However, no significant candidates were identified: see \citet{Udall:2019wtd} for further discussion. 
The most stringent upper limit on the local IMBH merger rate from O1 and O2 is $0.20\,$Gpc$^{-3}$\,yr$^{-1}$ (in comoving units, 90\% confidence level), for binaries with equal component masses $m_1=m_2=100\,\Msun$ \citep{2019PhRvD.100f4064A}. 
Other groups have also searched LIGO-Virgo open data \citep{O1Data,O2Data} for possible IMBH events \citep{Zackay:2019btq,Nitz:2019hdf}. 
The O3 run started in April 2019 with significantly increased sensitivities for all three Advanced detectors compared to O1 and O2 \citep{2019PhRvL.123w1107T,2019PhRvL.123w1108A}; here we consider the implications of \ThisEvent, detected in the first half of the run, O3a (1 April through 1 October, 2019). 

\subsection{Astrophysics of IMBHs}

Observational evidence for IMBHs, usually defined as BHs with mass $10^2-10^5\,\Msun$ (see e.g.,\ \citealt{2004cbhg.symp...37V,2004IJMPD..13....1M}) has long been sought. IMBHs bridge the gap between stellar BHs and supermassive BHs (SMBHs) and might be the missing link to explain the formation of SMBHs \citep{2010A&ARv..18..279V,2019arXiv191109678G}.  IMBHs are predicted to form in the early Universe via direct collapse of very massive population~III stars (${\gtrsim}\,230\,\Msun$; e.g.,\ \citealt{2001ApJ...550..372F,2003ApJ...591..288H,2017MNRAS.470.4739S}) or through collapse of low-angular-momentum gas clouds (e.g.,\ \citealt{1994ApJ...432...52L,2003ApJ...596...34B,2006MNRAS.371.1813L,2006MNRAS.370..289B}), perhaps passing through a quasi-star phase (e.g.,\ \citealt{2010MNRAS.402..673B,2011MNRAS.414.2751B}). 
In stellar clusters, IMBHs are predicted to form via dynamical channels such as runaway collisions \citep{2002ApJ...576..899P, 2004Natur.428..724P, AtakanGurkan:2003hm} and hierarchical mergers of smaller BHs \citep{2002MNRAS.330..232C,2006ApJ...637..937O,2015MNRAS.454.3150G}, especially in metal-poor star clusters \citep{2016MNRAS.459.3432M}.  
However, there is no conclusive observational confirmation of IMBHs in globular clusters and other massive star clusters \citep{2002AJ....124.3270G,2005ApJ...634.1093G, 2008ApJ...676.1008N, 2010ApJ...710.1032A,2010ApJ...710.1063V, 2011A&A...533A..36L, 2012ApJ...755L...1M, 2012ApJ...753..103N,2012ApJ...750L..27S, 2013A&A...552A..49L,2013ApJ...769..107L, 2017MNRAS.468.2114P,2017Natur.545..510K, 2018NatAs...2..656L,2018ApJ...862...16T,2019MNRAS.488.5340B,2019MNRAS.482.4713Z,2019ApJ...875....1M}. 

Several ultra-luminous X-ray sources, defined as those with a total luminosity, assumed isotropic, of ${\gtrsim}\,10^{39}$\,erg\,s$^{-1}$, 
have been studied as IMBH candidates \citep{2001MNRAS.321L..29K,2001ApJ...547L..25M,2003ApJ...586L..61S,2004cbhg.symp...37V,2004IJMPD..13....1M,2011NewAR..55..166F,2012MNRAS.423.1154S,2015MNRAS.448.1893M,2015ApJ...812L..34W,2017IJMPD..2630021M,2017ARA&A..55..303K, 2020arXiv200204618L}, but only a few still support evidence for IMBHs. HLX-1 is possibly the strongest IMBH candidate from electromagnetic data  \citep{2009Natur.460...73F,2009ApJ...705L.109G,2011ApJ...743....6S,2012Sci...337..554W, 2015MNRAS.446.3268C,2012MNRAS.420.3599S}, pointing to an IMBH mass ${\sim}0.3-30\times 10^4\,\Msun$. 

Several IMBH candidates lie at the centers of dwarf galaxies and are associated with low-luminosity active-galactic nuclei (AGNs; \citealt{1989ApJ...342L..11F,2003ApJ...588L..13F,2004ApJ...607...90B,2004ApJ...610..722G,2005ApJ...619L.151B,2007ApJ...670...92G,2010ApJ...714..713S,2012ApJ...755..167D,2013ApJ...775..116R,2015ApJ...809L..14B,2015ApJ...809..101D,2016ApJ...829...57B,2016ApJ...817...20M,2017ApJ...836...20B,2018MNRAS.478.2576M}). Their estimated masses are close to (or above) the upper edge of the IMBH mass range. 
The final mass of \ThisEvent is close to the lower end of the IMBH mass range, in an apparent BH desert covering the mass range ${\sim}10^2-10^3\,\Msun$.  Moreover, this final mass is the first confirmation that IMBHs can form through the merger of two less massive BHs.

\subsection{Pair instability mass gap}

The mass of the primary component of \ThisEvent falls within the range where PI is expected to suppress BH formation. PI develops in a star when the effective production of electron--positron pairs in the stellar core softens the equation of state, removing pressure support \citep{2007Natur.450..390W}. This leads to a contraction of the core, raising the internal temperature up to the ignition of oxygen or silicon, and the star becomes unstable.  PI is expected to develop in stars with helium core mass $\gtrsim 32\,\Msun$. For helium cores $32\lesssim M_{\rm He}/\Msun \lesssim 64$, this instability manifests as pulsational pair instability (PPI): the star undergoes a number of oscillations that eject material and remove the stellar envelope, bringing the star back to a stable configuration after the resulting mass loss \citep{1967PhRvL..18..379B,2007Natur.450..390W,2014ApJ...792...28C,2016MNRAS.457..351Y}. After PPI, the star ends its life with a core-collapse supernova or with direct collapse, leaving a compact object less massive than 
expected in the absence of PPI \citep{2017ApJ...836..244W,2019ApJ...878...49W}. For helium cores $64\lesssim M_{\rm He}/\Msun\lesssim 135$, PI leads to a complete disruption of the star, leaving no compact object, while for even larger helium cores PI drives a direct collapse to a BH \citep{1964ApJS....9..201F, 1983A&A...119...61O,1984ApJ...280..825B, 2003ApJ...591..288H, 2007Natur.450..390W}.

The combined effect of PI and PPI is expected to carve a mass gap in the BH mass function, with lower boundary ${\sim}40-65\,\Msun$ and upper boundary $\gtrsim 120\,\Msun$ \citep{2003ApJ...591..288H,2016A&A...594A..97B,2017MNRAS.470.4739S,2017ApJ...836..244W,2018MNRAS.474.2959G,2019ApJ...878...49W}. The boundaries of the mass gap are highly uncertain because they depend on stellar evolution and on our understanding of core-collapse, PPI and PI supernovae \citep{2019ApJ...882..121S,2019ApJ...882...36M,2019ApJ...887...53F,2020ApJ...888...76M}. 
Several formation channels might populate the mass gap. Below, we will review these channels and attempt to interpret the astrophysical origin of \ThisEvent in this context and to put constraints on different scenarios.

\subsection{Outline of the paper}

We describe the detection of \ThisEvent in a companion paper \citep{GW190521-Discovery}, where we detail the circumstances of the observation and the detection significance using three different search pipelines.  
The search results are consistent with a coherent astrophysical signal and inconsistent with an instrumental noise origin for the event. 
In addition, the strain data are consistent with GW emission from the coalescence of a quasi-circular compact binary system. 

In this paper, we begin by assuming that 
the source is indeed the coalescence of such a 
binary. In Section~2, we give further details about the Bayesian parameter estimation procedure, and the posterior probability distributions that provide estimates of the source's intrinsic and extrinsic parameters. 
We quantify the evidence for orbital precession due to in-plane component spins, and the evidence for the presence of higher-order multipoles beyond the dominant $\ell=2, m=2$ mode in the data.  

In Section~3, we discuss the consistency of the observed signal with 
the coalescence of a quasi-circular compact binary system. We test the consistency of the residual data, after subtraction of the best-fitting signal, with detector noise, and the consistency of the merger and ringdown portions of the signal with expectations from waveform models derived from GR. 

In Section~4, we present an estimate of the rate per comoving volume for merger events similar to \ThisEvent. In Section~5, we consider the astrophysical implications of the observation, discussing uncertainties on the PI mass gap and proposed astrophysical channels that might populate the mass gap, including hierarchical merger in stellar cluster environments and stellar mergers.  In Section~6, we discuss alternative scenarios for the source of \ThisEvent, including a strongly gravitationally lensed merger, an eccentric BBH, or a primordial BBH; we also exclude a cosmic string cusp or kink or a core-collapse supernova as possible sources due to mismatch with the GW data. 
Finally, we summarize our observations and consider future prospects in Section~7.

\begin{table*}[t!]
\caption{Source properties for \ThisEvent: \editNew{median values with 90\% credible intervals} that include statistical errors.}
\begin{ruledtabular}
\begin{tabular}{l l l l }
Waveform Model  & NRSur~PHM  & Phenom~PHM  & SEOBNR~PHM   \\
\hline
Primary BH mass $m_1~(\Msun)$ & \mOne &  $90^{+23}_{-16}$ & $99^{+42}_{-19}$ \\
\rule{0pt}{3ex}%
Secondary BH mass $m_2~(\Msun)$& \mTwo & $65^{+16}_{-18}$ & $71^{+21}_{-28}$ \\
\rule{0pt}{3ex}%
Total BBH mass $M~(\Msun)$ & \mTotal & $154^{+25}_{-16}$ & $170^{+36}_{-23}$  \\
\rule{0pt}{3ex}%
Binary chirp mass $\mathcal{M}~(\Msun)$ & \mChirp & $65^{+11}_{-7}$ & $71^{+15}_{-10}$  \\
\rule{0pt}{3ex}%
Mass-ratio $q=m_2/m_1$ & \massRatio & $0.73^{+0.24}_{-0.29}$ & $0.74^{+0.23}_{-0.42}$  \\

\hline
\rule{0pt}{3ex}%
Primary BH spin $\chi_1$ & \aOne  & $0.65^{+0.32}_{-0.57}$ & $0.80^{+0.18}_{-0.58}$ \\
\rule{0pt}{3ex}%
Secondary BH spin $\chi_2$ & \aTwo  & $0.53^{+0.42}_{-0.48}$  & $0.54^{+0.41}_{-0.48}$  \\
\rule{0pt}{3ex}%
Primary BH spin tilt angle $\theta_{LS_1}~(\mathrm{deg})$ & \thOne  & $80^{+64}_{-49}$  & $81^{+49}_{-45}$ \\
\rule{0pt}{3ex}%
Secondary BH spin tilt angle $\theta_{LS_2}~(\mathrm{deg})$ & \thTwo  & $88^{+63}_{-58}$  & $93^{+61}_{-60}$  \\
\rule{0pt}{3ex}%
Effective inspiral spin parameter $\chi_\mathrm{eff}$ & \chiEff & $0.06^{+0.31}_{-0.39}$ & $0.06^{+0.34}_{-0.35}$\\
\rule{0pt}{3ex}%
Effective precession spin parameter $\chi_\mathrm{p}$ & \chiP & $0.60^{+0.33}_{-0.44}$ & $0.74^{+0.21}_{-0.40}$\\

\hline
\rule{0pt}{3ex}%
Remnant BH mass $M_\mathrm{f}~(\Msun)$ & \mFinal & $147^{+23}_{-15}$ & $162^{+35}_{-22}$ \\
\rule{0pt}{3ex}%
Remnant BH spin $\chi_\mathrm{f}$ & \aFinal & $0.72^{+0.11}_{-0.15}$ & $0.74^{+0.12}_{-0.14}$ \\
\rule{0pt}{3ex}%
Radiated energy $E_\mathrm{rad}~(\Msun c^2)$ & \mTotalMinusMfinal & $7.2^{+2.7}_{-2.2}$ & $7.8^{+2.8}_{-2.3}$ \\
\rule{0pt}{3ex}%
Peak Luminosity $\ell_\mathrm{peak}~(\mathrm{erg}~\mathrm{s}^{-1})$ & \lPeak $\times 10^{56}$ & $3.5^{+0.7}_{-1.1} \times 10^{56}$ & $3.5^{+0.8}_{-1.4} \times 10^{56}$  \\
\hline
\rule{0pt}{3ex}%
Luminosity distance $D_\mathrm{L}~(\mathrm{Gpc})$ & \LuminosityDistance  & $4.6^{+1.6}_{-1.6}$ & $4.0^{+2.0}_{-1.8}$  \\
\rule{0pt}{3ex}%
Source redshift $z$ & \redshift & $0.73^{+0.20}_{-0.22}$& $0.64^{+0.25}_{-0.26}$  \\
\rule{0pt}{3ex}%
Sky localization $\Delta{\Omega}~(\mathrm{deg}^2)$ & $\skyareaLalNRSur$ & $\skyareaLalIMRPhenom$ & $\skyareaRIFTSEOBNR$

\end{tabular}
\end{ruledtabular}
\label{tab:parameters}
\end{table*}

\section{Source Properties}
\label{sec:PE}

Under the assumption that \ThisEvent is a quasi-circular BBH coalescence, the intrinsic parameters of the source are fully described by the masses, $m_{1}$ and $m_{2}$, and the spin vectors, ${\bf S}_{1}$ and ${\bf S}_{2}$, of the two BHs. 
We use the convention that $m_{1} \geq m_{2}$ and the mass-ratio $q=m_2/m_1 \leq 1$.
The dimensionless spin magnitudes, $\chi_i = |c\,{\bf S}_i/(G m_i^2)|$, are assumed to be constant throughout the inspiral, while the spin orientations relative to the orbital angular momentum axis, $\theta_{LS_i}=\arccos(\hat{\bf {L}}\cdot{\hat{\bf S}_i)} \in [0, 180^\circ]$, evolve over the duration of the signal. 
The spin orientations must therefore be parameterized at some fiducial time which, for this work, is when the signal has a GW frequency $f_0 = 11$\,Hz.
The remnant BH produced from post merger is described by its mass $M_\mathrm{f}$ and spin magnitude $\chi_\mathrm{f}$. 
We also estimate the BH recoil velocity $v_\mathrm{f} = |\bf{v}_\mathrm{f}|$ relative to the center of mass of the binary. 

 \editNew{As discussed in Section 2.4 below, the source is estimated to be cosmologically distant. The masses measured in the frame of LIGO and Virgo detectors are therefore redshifted by a factor $(1+z)$ and are denoted with a superscript $det$ so that $m^\mathrm{det} = (1+z)\,{}m$, where $m$ is the source frame mass. GWs directly encode the luminosity distance to the source $D_\mathrm{L}$, which in turn depends on the inclination angle of the binary orbit with respect to the line of sight (Section 2.4).  To make inferences about the source frame masses we must therefore convert the distance measurement to a redshift. The statistical uncertainty associated with estimation of the source frame masses is increased relative to that of the detector frame masses due to these dependencies.} From the inferred posterior distribution of $D_\mathrm{L}$ we compute redshift assuming a Planck 2015 $\Lambda$CDM cosmology with Hubble parameter $H_0=67.9$\,km\,s$^{-1}$\,Mpc$^{-1}$ \citep{Ade:2015xua} \editNew{(and we address the effect of taking a larger value of the Hubble parameter in Section 2.2)}. Unless stated otherwise, mass measurements are quoted in the source frame of the binary.

\subsection{Method and Signal Models}
\label{sec:PE-method}
To infer the source properties of \ThisEvent, we analyzed \seglen\ seconds of data in the LIGO and Virgo detectors around the time of the detection. 
The data are downsampled from 16384\,Hz to \samplerate\,Hz, as we expect no signal power above several hundred Hz due to the total mass of \ThisEvent.
The parameter estimation analysis is done with two independently-developed coherent Bayesian inference pipelines -- \texttt{LALInference} \citep{PhysRevD.91.042003} and \texttt{RIFT} \citep{Lange:2018pyp, Wysocki_2019}, which produce consistent results for the inferred source parameters. 
Both parameter estimation algorithms assume stationary Gaussian noise characterized by the power spectral density (PSD) which is inferred from the data by the \texttt{BayesLine} algorithm \citep{PhysRevD.91.084034}.
We compute the event likelihood in the frequency domain, integrating over the frequency band $\flow - \fhigh$\,Hz.

We used three distinct GW signal models of BBH coalescence in our analysis: \texttt{NRSur7dq4} (NRSur~PHM), a surrogate waveform model built by directly interpolating 
NR solutions \citep{PhysRevResearch.1.033015}; \texttt{IMRPhenomPv3HM} (Phenom~PHM), an inspiral-merger-ringdown waveform model 
that uses phenomenological frequency-domain fits combining post-Newtonian calculations of the GW phase and amplitude 
\citep{Blanchet:1995ez,Damour:2001bu,2005PhRvD..71l4004B,Arun:2008kb,Blanchet:2013haa} with tuning to NR solutions \citep{PhysRevD.101.024056}; \texttt{SEOBNRv4PHM} (SEOBNR~PHM), an inspiral-merger-ringdown waveform model that is based on the effective-one-body formalism \citep{Buonanno:1998gg,Buonanno:2000ef} and calibrated to NR \citep{SEOBNRV4PHM}.

These three waveform models employ different approaches to reproduce the predictions from analytical and numerical relativity; we expect to see differences in the parameter estimation from these three models due to those different approaches, and we can interpret those differences as a form of systematic error associated with the modeling.  \editNew{Note that the effects of the astrophysical environment, such as the presence of gas, on the GW waveform is expected to be negligible \citep{PhysRevLett.119.171103} in the \editNew{late} stages of inspiral, merger and ringdown that we observe.}

NRSur~PHM is constructed based on NR simulations with component spins that are not constrained to be aligned with the orbital axis, thus including the effects of spin-orbit precession. The model covers dimensionless spin magnitudes $\chi_i \leq 0.8$ and mass ratios $q = m_2/m_1 \geq 1/4$. 
It includes all $(l,|m|)$-multipoles of the gravitational radiation \citep{Blanchet:1996pi,Kidder:2007rt,Blanchet:2008je,Mishra:2016whh} up to and including $l = 4$. In the training parameter space, the waveform model has shown excellent agreement with NR simulations, with mismatches comparable to the numerical errors associated with the NR simulation. The model continues to agree with NR when extrapolating to mass ratios $q\geq 1/6$ \citep{PhysRevResearch.1.033015}. NRSur~PHM is directly trained with NR simulations and therefore only models the last ${\sim}20$ orbits of the inspiral, which is adequate for \ThisEvent because the signal is in the measurement band of the detectors for fewer cycles.

Phenom~PHM \citep{PhysRevD.100.024059} is an approximate higher-multipole aligned-spin waveform model that maps the subdominant radiative moments $(l,|m|) = (2,1),(3,2),(3,3),(4,3),(4,4)$ to the dominant $(2,2)$ mode \citep{London:2017bcn}.  \editNew{Multipoles are defined in the co-precessing frame where the binary approximates a system with aligned spins \citep{Schmidt:2010it,Pekowsky:2013ska}, and are then transformed by a time-dependent rotation to model the harmonic modes of a precessing binary in the inertial frame \citep{Schmidt:2012rh,Hannam:2013oca}, using a double-spin model of spin-orbit precession during the inspiral \citep{Chatziioannou:2017tdw}}. After this precession ``twisting", all $l \leq 4$ modes will be non-zero in the inertial frame.  The non-precessing, dominant multipole of the radiation is tuned to spin-aligned NR simulations in the parameter space of spin magnitudes $\chi_i \leq 0.85$ and mass ratios $q \geq 1/18$ \citep{Husa:2015iqa,Khan:2015jqa}. The subdominant multipoles and precessional effects in Phenom~PHM, however, have not been calibrated to NR, and the model does not include spherical-spheroidal mode-mixing effects which can significantly impact some of the higher multipoles \citep{Kelly:2012nd}.

SEOBNR~PHM is based on the dynamics of spinning, non-precessing BBHs in the effective-one-body formalism, calibrated to NR simulations and results from BH perturbation theory \citep{PhysRevD.95.044028}. The model includes the non-precessing $(l,|m|) = (2,1),(3,3),(4,4),(5,5)$ multipoles in addition to the dominant $(2,2)$ multipole mode. The individual modes are calibrated to waveforms from NR and BH perturbation theory \citep{Cotesta:2018fcv}, covering the parameter space of mass ratios $q\geq 1/10$ and effective inspiral spin parameter $\chi_{\rm eff} \in [-0.7, 0.85]$. Effects from the precessing orbital plane are modeled through a suitable rotation of the non-precessing inspiral-plunge multipoles from the co-precessing frame to the inertial frame (without recalibration to NR), with a direct attachment of merger-ringdown modes in the co-precessing frame \citep{PhysRevD.95.024010,SEOBNRV4PHM}. After the precession twisting, all $l \leq 5$ modes will be non-zero in the inertial frame.  

Although not directly fitted to NR simulations of precessing binaries, both Phenom~PHM and SEOBNR~PHM have been validated through a comparison with a large set of such NR waveforms \citep{PhysRevD.101.024056, SEOBNRV4PHM}.  \editNew{The three models described above are tuned to different NR solutions. Comparisons between different NR codes find agreement on the level of accuracy of the individual codes~\cite{Hannam:2009hh, Hinder:2013oqa, Lovelace:2016uwp}.  The agreement between different NR codes is sufficiently good~\cite{Purrer:2019jcp} to avoid systematic biases at the SNR of \ThisEvent. However, The three models are constructed in sufficiently different ways that it is useful to compare the results of parameter estimation from each of them.}

The NRSur~PHM waveform model is 
most faithful to NR simulations in the parameter range relevant for \ThisEvent \citep{PhysRevResearch.1.033015}. Therefore, the inferred source parameters quoted in this paper and in \citep{GW190521-Discovery} were obtained with the NRSur~PHM model unless otherwise noted. 
In this section we also present results from the Phenom~PHM and SEOBNR~PHM models to check for systematic differences between waveform models. As shown below, differences in results between waveform models are of the same order or smaller than the statistical error, and do not impact the astrophysical interpretation of \ThisEvent discussed in this paper.

\begin{figure*}[t!]
\begin{center}
    \includegraphics[width=\columnwidth]{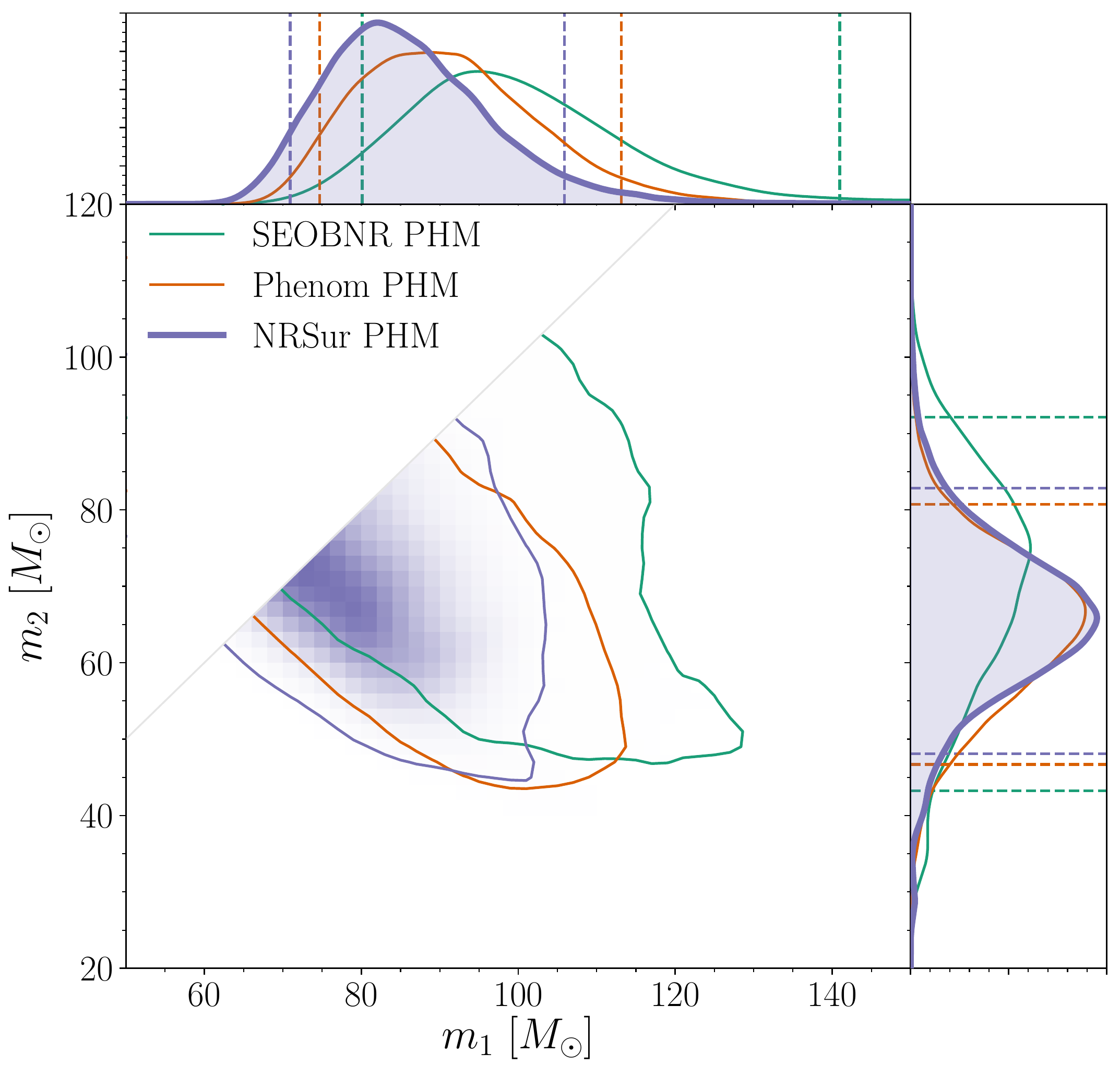}
	\includegraphics[width=\columnwidth]{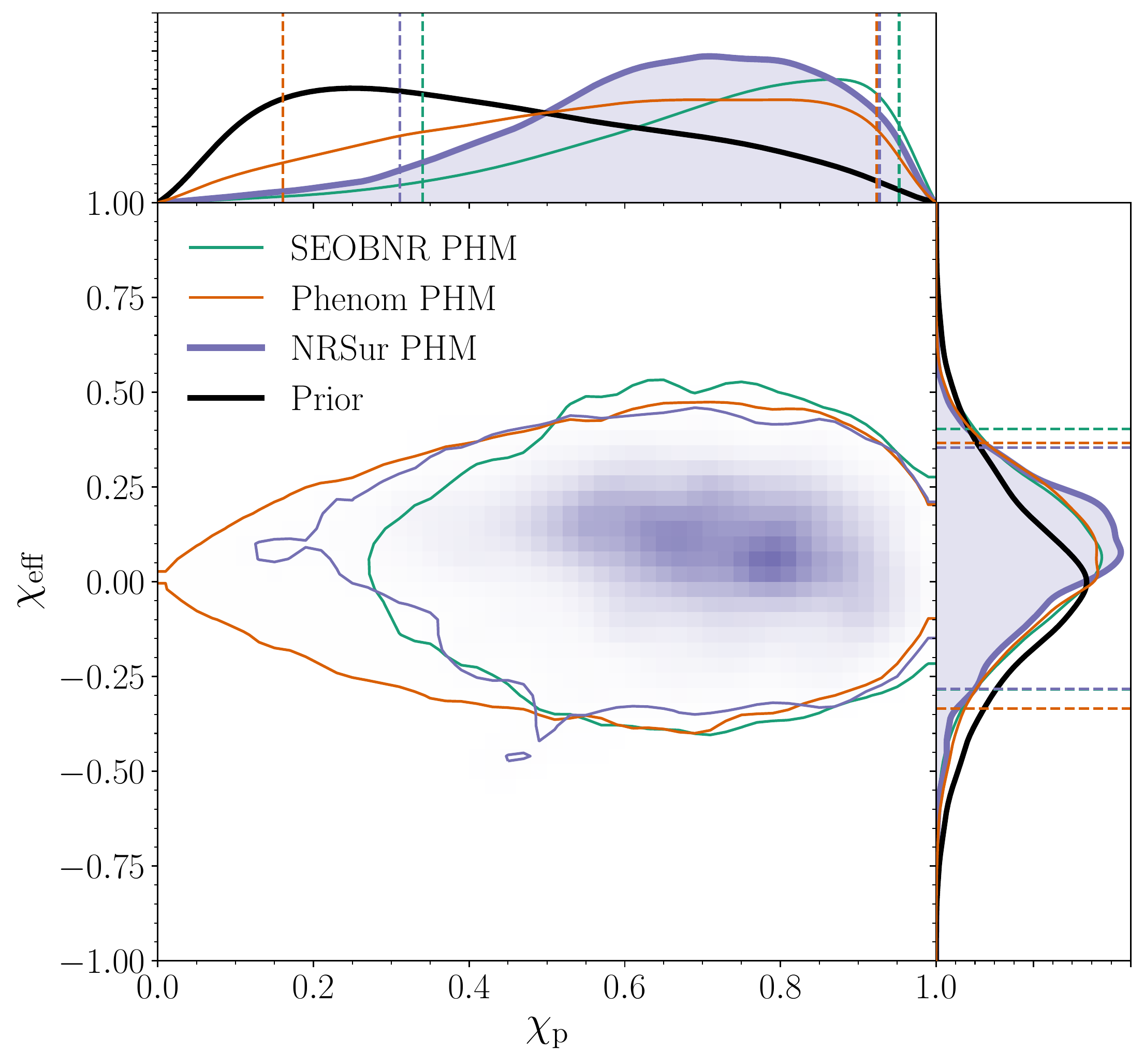}
   \caption{Posterior distributions on the individual source-frame masses (left) and effective spin parameters (right) according to the three waveform models employed. The one-dimensional distributions include the posteriors for the three waveform models, and the dashed lines mark their 90\% credible interval. The two-dimensional plot shows the 90\% credible regions for each waveform model, with lighter-blue shading showing the posterior distribution for the NRSur~PHM model. The black lines in the right panel show the prior distributions.}
   \label{fig:masses}
   \end{center}
\end{figure*}

We choose priors that are uniform on the component masses in the detector frame from $\ComponentMassPrior$. 
We further restrict the mass priors such that the total mass must be greater than $200\,\Msun$, and the chirp mass 
to be between 70 and $150\,\Msun$, both in the detector frame.
In all cases, we verify that the posterior distributions do not have support at the boundaries of the priors.
The distance prior scales as $D_\mathrm{L}^2$ ({\it i.e.}, differential in Euclidean volume) up to a maximum of $\LuminosityDistancePrior$\,Gpc. 
 \editNew{We have checked that a prior that is differential in comoving volume (with Planck 2015 cosmology) makes negligible ($< 1$\%) difference in the posterior medians for all source parameters.}
For the BH spins, we adopt a uniform prior for the magnitude in the dimensionless parameters $\chi_i \in [0,\,0.99]$ and their orientation angles chosen to be uniform on the surface of the unit sphere.
We adopt a uniform prior in the cosine of the inclination angle between the binary angular momentum and the line of sight, $\theta_\mathrm{JN}$.
The prior on the sky location (right ascension and declination) is chosen to be uniform on the surface of the unit sphere.

The parameters of the remnant BH formed after the merger, its mass
$M_\mathrm{f}$, dimensionless spin $\chi_\mathrm{f}$, and recoil kick velocity
$v_\mathrm{f}$ are inferred by applying fits calibrated to NR to the posterior 
distributions of the binary's initial masses and spins. For the
posteriors from the Phenom~PHM and SEOBNR~PHM waveform models, we used the same
$M_\mathrm{f}$ and $\chi_\mathrm{f}$ fits that are implemented internally in these
models: for Phenom~PHM, the fits from \cite{Husa:2015iqa} with corrections for
precession from \citet{PhenomPv2_tech_doc}, and for SEOBNR~PHM, the fits from
\citet{SEOBNRV4PHM} and from \citet{2016ApJ...825L..19H} applied to the spins
evolved using the waveform model's dynamics as described in
\citet{SEOBNRV4PHM}. For the NRSur~PHM posterior, we applied the related
surrogate remnant fit of \citet{PhysRevResearch.1.033015} for $M_\mathrm{f}$,
$\chi_\mathrm{f}$ and $v_\mathrm{f}$. 
Applying the $M_\mathrm{f}$ and $\chi_\mathrm{f}$ surrogate fits to the 
posteriors from the Phenom~PHM and SEOBNR~PHM waveform models, \editNew{and} using the 
average of the fits from \citet{Healy:2016lce}, \citet{2016ApJ...825L..19H} and 
\citet{Jimenez-Forteza:2016oae} after applying corrections for precession 
\citep{spinfit-T1600168, GW170104}, \editNew{both} yield consistent results. The surrogate 
$v_\mathrm{f}$ was only tested for NRSur~PHM \citep{Varma:2020kick}, therefore we
do not apply it to Phenom~PHM and SEOBNR~PHM.
The peak luminosity is also inferred using fits calibrated to numerical relativity \citep{Healy:2016lce,Keitel:2016krm}, while the energy radiated in the merger is given by $M - M_\mathrm{f}$.

\editNew{The key analysis elements described above, including parameter estimation sampling algorithms, PSD estimates, and waveform models, all potentially introduce systematic uncertainties. Different choices for these elements can affect the results but in most cases these changes are significantly smaller than the statistical uncertainties. Below, we highlight the more significant differences in the results associated with waveform models.}

\subsection{Primary and Secondary Black Hole Components}
\label{sec:PE_components}

In Table~\ref{tab:parameters} we summarize the source properties of \ThisEvent. Results are quoted as the median and symmetric 90\% credible interval of the marginalized posterior distributions for each parameter, and for each of the three GW signal models. The measurements are marginalized over uncertainty in the data calibration.
In the rest of this paper we quote source properties derived using NRSur~PHM,  
unless explicitly stated otherwise. 

\paragraph{Masses} The estimated mass posterior distributions is shown in Fig.~\ref{fig:masses} (left) for the three GW signal models. The primary BH mass of \ThisEvent is {$m_1 = \mOne\,\Msun$}, making it the highest-mass component BH known to date in GW astronomy. 
The mass of the secondary BH is inferred to be {$m_2 = \mTwo\,\Msun$}.
The primary BH of \ThisEvent is more massive (median value) than any \emph{remnant} BH reported in GWTC-1 except for GW170729 \citep{LIGOScientific:2018mvr}; the secondary BH of \ThisEvent is also more massive than any \emph{primary} BH in GWTC-1.

 \editNew{These source frame masses have been redshift-corrected, as discussed above, using a value of the Hubble parameter $H_0=67.9$ from Planck 2015. However, recent measurements of $H_0$ using nearby Cepheid distance standards obtain a precise value of $H_0=74.03\pm1.42$\,km\,s$^{-1}$\,Mpc$^{-1}$ \citep{Riess_2019},  9\%\ higher than the Planck value. Using this latter value along with the other cosmological parameters from Planck 2015 increases the median value of the redshift by 7\%\ and reduces the estimated source frame masses by 3\%. These shifts are significantly smaller than statistical or other systematic uncertainties, including those affecting the astrophysical interpretation discussed throughout this paper.}

While the low mass cutoff of the PI mass-gap is uncertain (see Section~\ref{sec:PI_uncertainty}), the primary BH of \ThisEvent offers strong evidence for the existence of BHs in the mass-gap. If the PI gap begins at $50\,\Msun$ $(65\,\Msun)$, we find that the primary BH has only a {\mOneProbLessThanFifty\% (\mOneProbLessThanSixtyFive\%)} probability of being below the mass-gap, while the secondary BH has {\mTwoProbLessThanFifty\% (\mTwoProbLessThanSixtyFive\%)} probability of also being below the mass-gap. 

The SEOBNR~PHM model supports a higher primary mass and more asymmetric mass ratio for \ThisEvent: within 90\% credible intervals, $m_1$ and $m_2$ can be as high as \editNew{$141\,\Msun$ and $92\,\Msun$} respectively, while support for the mass ratio extends down to \editNew{$q \sim 0.32$}. 
While the upper limit of the PI mass gap remains uncertain, adopting $120\,\Msun$ as the high mass end of the gap we find the probability that the primary BH of \ThisEvent is beyond the gap of \editNew{$\mOneProbMax\%$} when using the SEOBNR~PHM model.  The corresponding probabilities using the NRSur~PHM and Phenom~PHM models are \editNew{$0.9\%$ and $2.3\%$}, respectively.

The probability that at least one of the black holes in \ThisEvent is in the range $65 - 120\,\Msun$ is \editNew{\mOneTwoProbInGap\%}, using the NRSur~PHM model. The corresponding probabilities using the Phenom~PHM  and SEOBNR~PHM models are \editNew{98.0\%\ and 90.2\%}, respectively.

We measure the total binary mass of \ThisEvent to be \editNew{$M= \mTotal\,\Msun$} making it the highest-mass binary observed via GWs to date. The binary chirp mass is \editNew{$\mathcal{M} = \mChirp\,\Msun$}, a factor ${\sim}2$ times higher than the first BBH detection, GW150914 \citep{GW150914:PE, LIGOScientific:2018mvr}. \ThisEvent is consistent with a nearly equal mass binary with mass ratio \editNew{$q = m_2/m_1 = \massRatio$} (90\% credible interval).

In the detector frame, the measured masses are \editNew{$m_1^{det} = \mOneDet\,\Msun$, $m_2^{det} =  \mTwoDet\,\Msun$, $M^{det} = \mTotalDet\,\Msun$, and $\mathcal{M}^{det} =  \mChirpDet\,\Msun$}, using the NRSur~PHM model. These results are very nearly the same for all three models.

\begin{figure*}[t!]
\begin{center}
\includegraphics[scale=0.42, trim = {0 0 0 0 }]{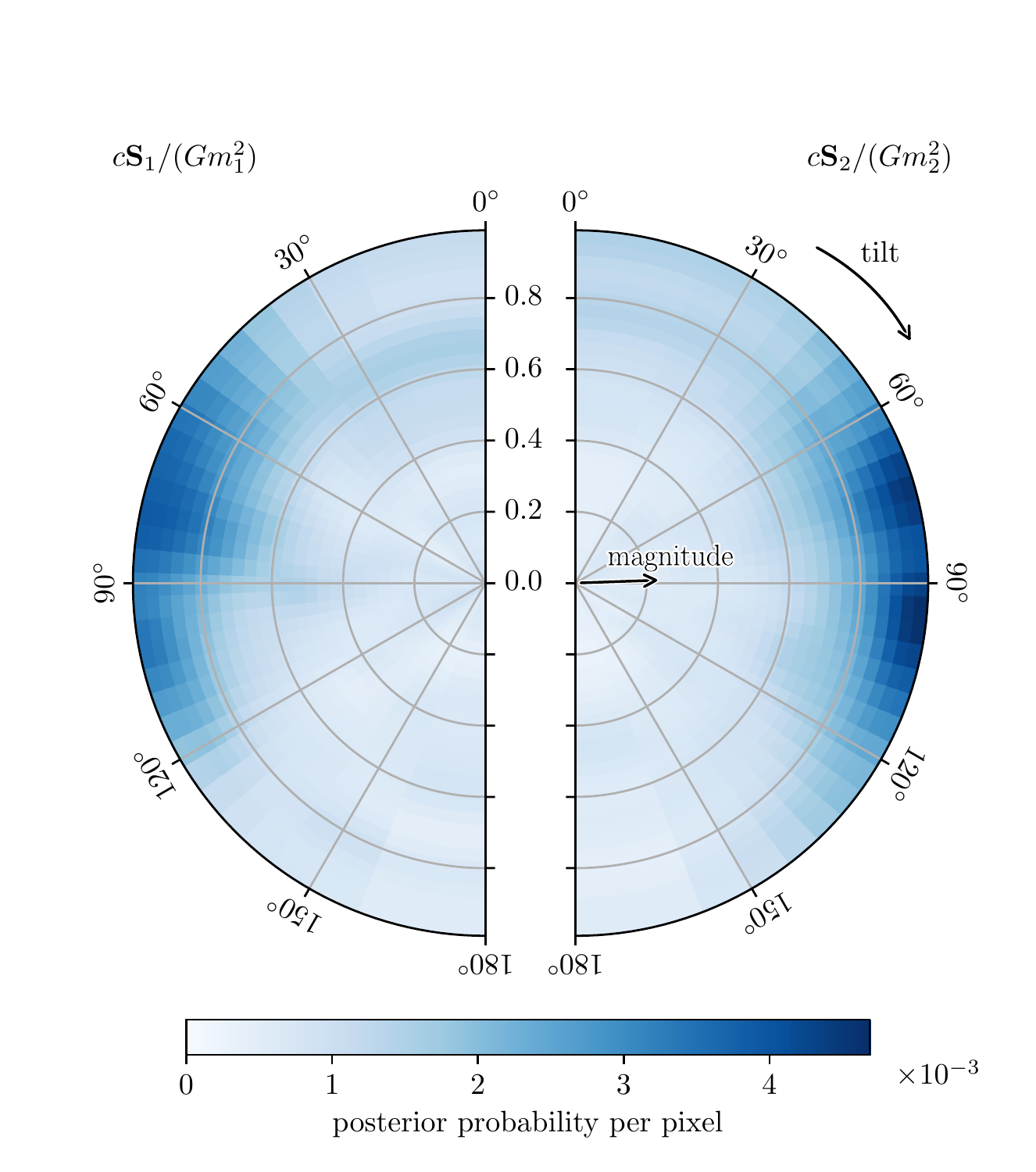}
\includegraphics[scale=0.42, trim = {0 0 0 0 }]{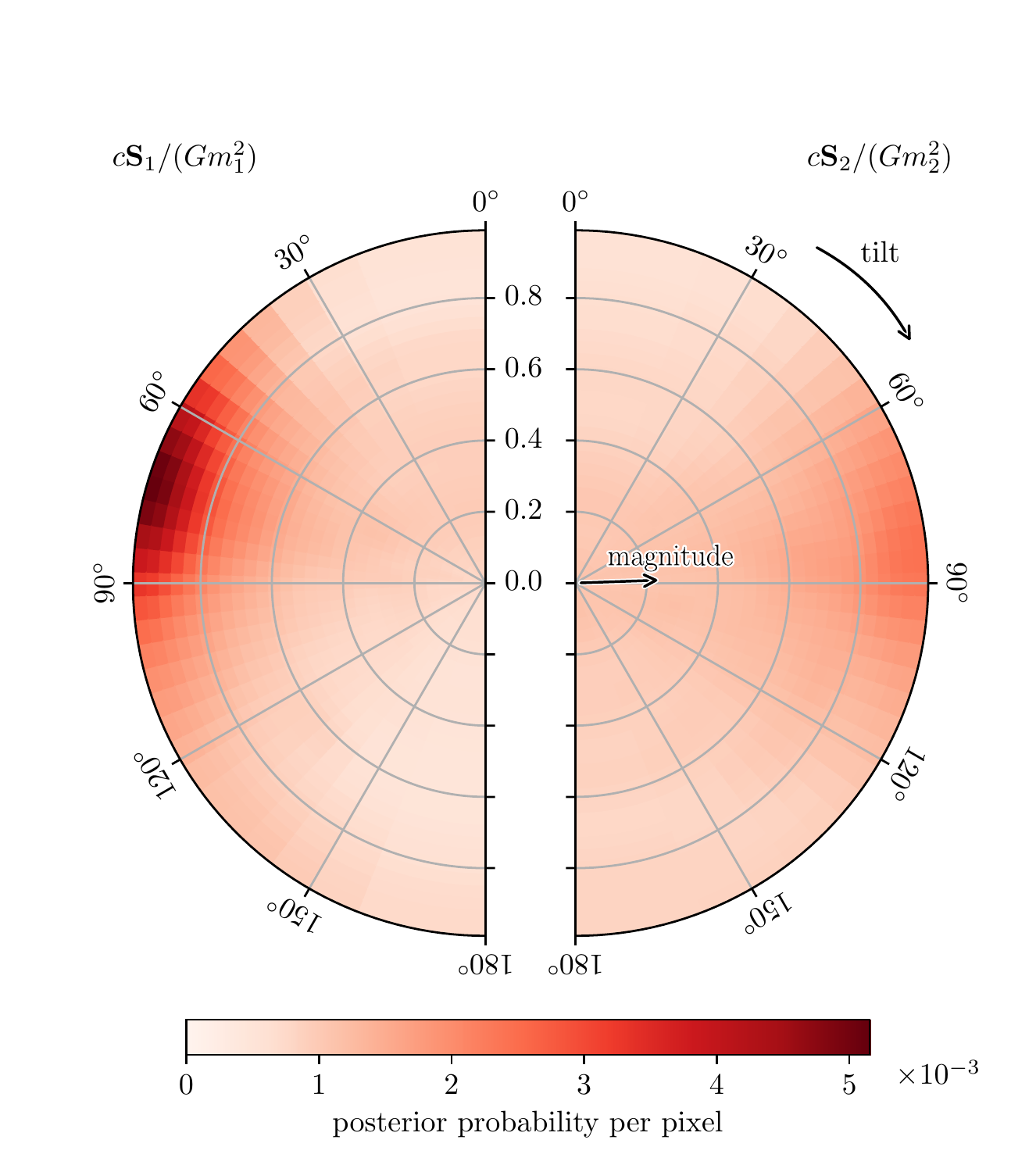}
\includegraphics[scale=0.42, trim = {0 0 0 0 }]{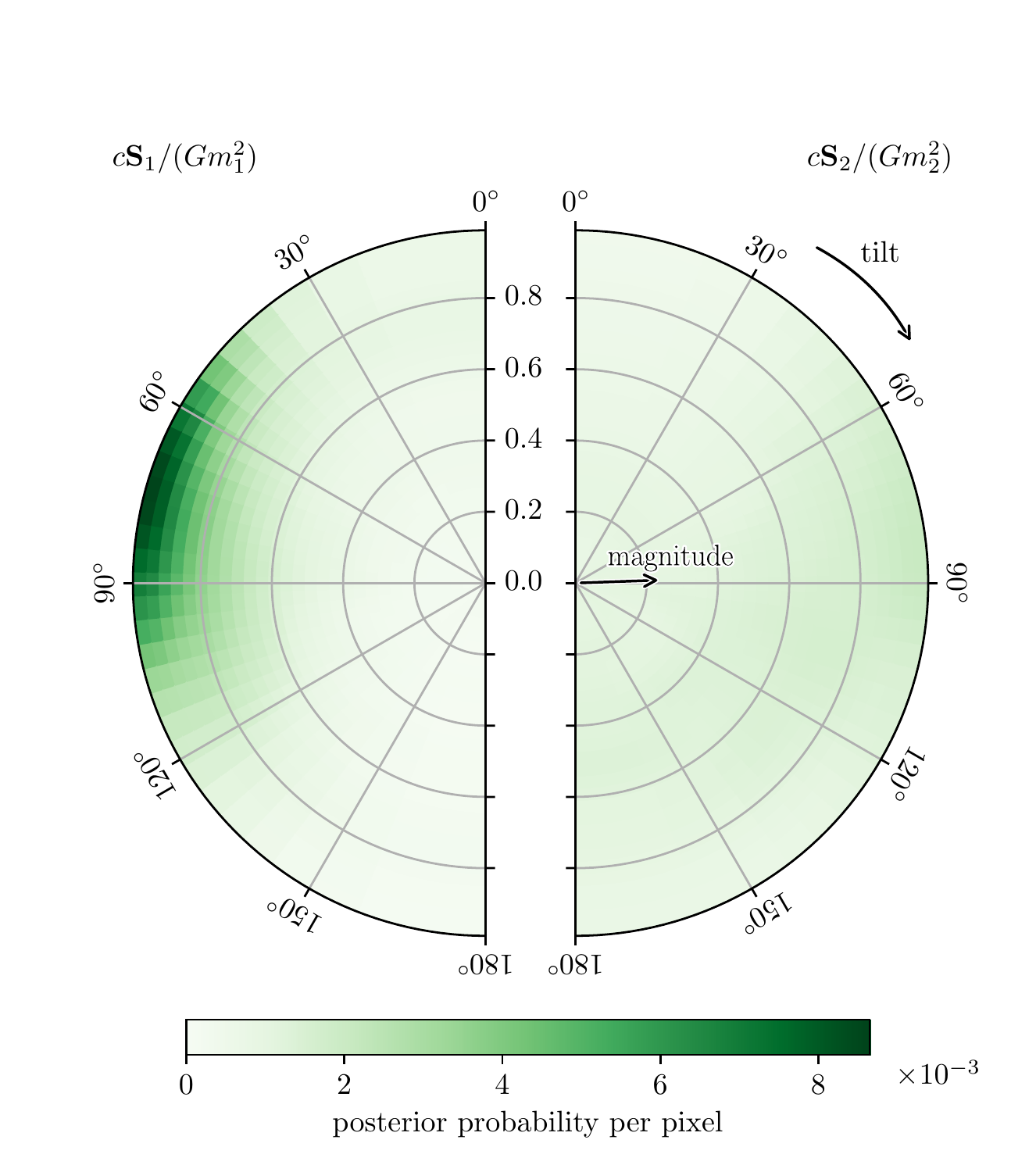}
  \caption{Posterior probabilities for the dimensionless component spins, $c\,{\bf S}_1/(Gm_1^2)$ and $c\,{\bf S}_2/(Gm_2^2)$, relative to the orbital angular momentum axis $\hat{\bf L}$. Shown here for the three waveform models (left to right: NRSur~PHM, Phenom~PHM, and SEOBNR~PHM).  The tilt angles are $0^\circ$ for spins aligned with the orbital angular momentum and $180^\circ$ for spins anti-aligned.  Probabilities are marginalized over the azimuthal angles. The pixels have equal prior probability, being equally spaced in the spin magnitudes and the cosines of tilt angles. The spin orientations are defined at a fiducial GW frequency of 11\,Hz.  
  }
   \label{fig:spins}
   \end{center}
\end{figure*}

\paragraph{Spins} Due to its high total mass, \ThisEvent is the shortest duration signal (approximately 0.1\,s) recorded so far in the LIGO and Virgo detectors. With only around 4 cycles (2 orbits) in the frequency band $30-80$\,Hz 
\citep{GW190521-Discovery}, information about spin evolution during the coalescence 
is limited. 
Still, analyses of \ThisEvent indicate that GW signal models including effects of spin-orbit precession are mildly preferred over those that omit such effects (i.e., allow only spins aligned with the orbital axis), with a $\log_{10}$-Bayes factor of \editNew{$\logTenBFprecession$} for the NRSur~PHM model allowing generic BH spins vs.\ limiting the effects of spin to the aligned components. 

In the disk plots of Fig.~\ref{fig:spins}, we show constraints on the spins of the component BHs of \ThisEvent in terms of their dimensionless magnitudes $\chi_{1}$ and $\chi_2$ and polar angles (tilts) with respect to the orbital angular momentum, $\theta_{LS_1}$ and $\theta_{LS_2}$, defined at a fiducial GW frequency of 11\,Hz.  Median values from all three waveform models suggest in-plane spin components with high spin magnitudes for both the BHs. Within the 90\% credible intervals given in Table~\ref{tab:parameters}, however, the constraints on the dimensionless BH spin magnitudes remain uninformative. For our preferred model NRSur~PHM, the 90\% bounds on spin magnitude extend from $\chi_{1,2} \sim 0.1-0.9$. The constraints on the tilt angles of these spins are also relatively broad.

As for past GW observations, we present inferences on the spins of \ThisEvent using the parameters $\chi_\mathrm{eff}$ and $\chi_\mathrm{p}$ constructed from the mass and spin of the binary components. Here, $\chi_\mathrm{eff} = (m_1\chi_1\cos{\theta_1} + m_2\chi_2\cos{\theta_2})/(m_1 + m_2) \in [-1,~1]$ is the effective inspiral spin parameter \citep{PhysRevD.64.124013,Ajith:2009bn} which measures the mass-weighted net spin aligned with the orbital angular momentum axis $\hat{\bf{L}}$ and remains approximately constant throughout the inspiral \citep{PhysRevD.78.044021}. The effective precession spin parameter $\chi_\mathrm{p} \in [0,~1]$ \citep{Schmidt_2015} measures the spin components in the plane of the orbit, and therefore the strength of the spin-orbit precession in the binary.  
The inferred posterior distributions of $\chi_\mathrm{eff}$ and $\chi_\mathrm{p}$ are shown in Fig.~\ref{fig:masses} (right) for the three GW signal models. Our priors are uniform in the component spins of the binaries, which results in nontrivial priors for the effective spin parameters, shown as the black distribution in the figure. For all three waveform models the posterior distribution of $\chi_\mathrm{eff}$ is peaked close to $0$, similarly to the prior, while the $\chi_\mathrm{p}$ distribution is shifted towards higher values. 
While the bulk of the posterior on $\chi_\mathrm{p}$ suggests an in-plane component of the spins, which contributes to spin-induced precession, the broad distribution prevents a more conclusive finding.

\begin{figure}[tb!]
\begin{center}
\includegraphics[scale=0.38, trim = {0 0 0 0 }]{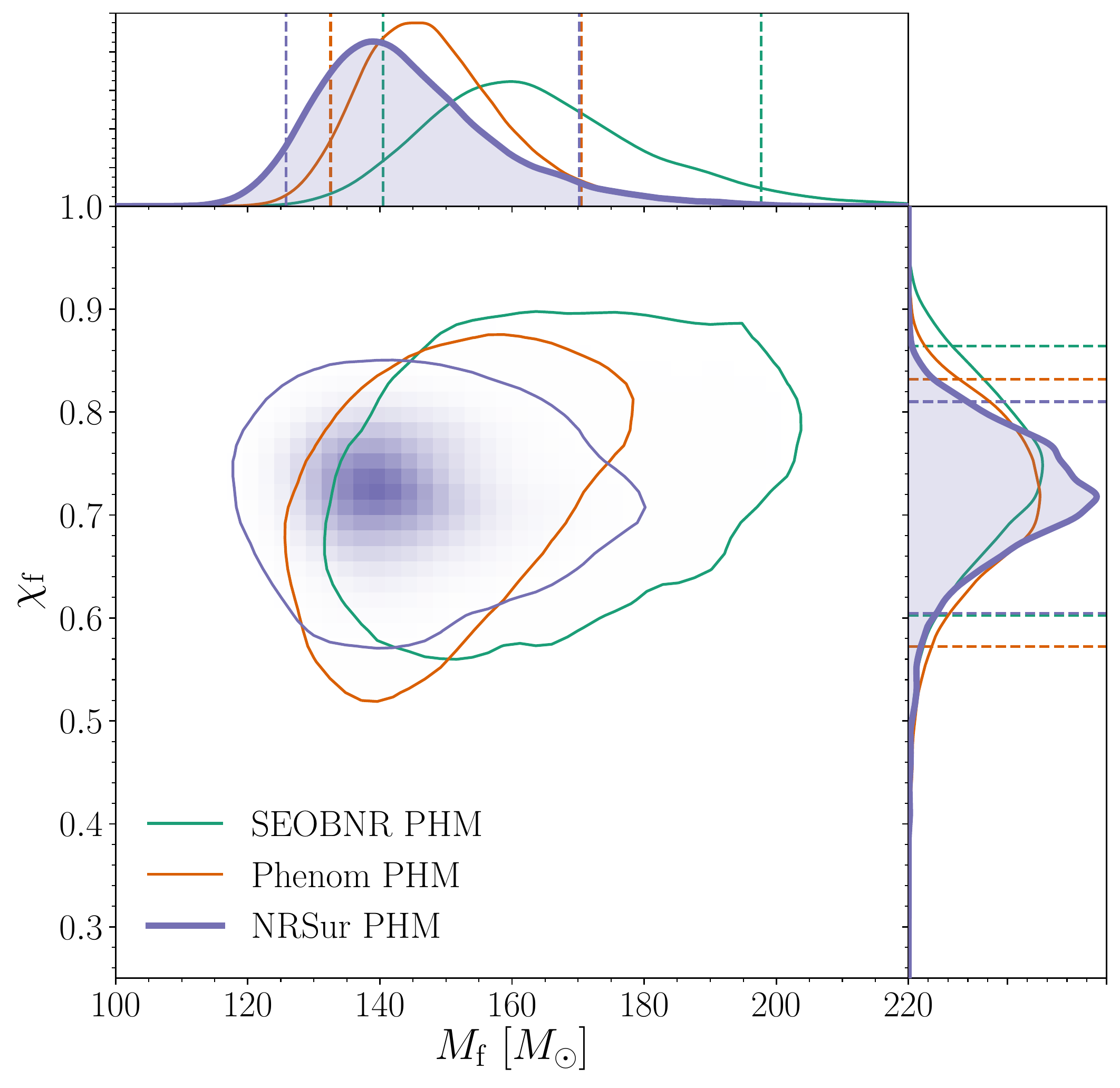}
   \caption{Posterior distributions of the mass ($M_\mathrm{f}$) and the dimensionless spin ($\chi_\mathrm{f}$) of the remnant BH according to the three waveform models employed. The one-dimensional distributions include the posteriors for the three waveform model, and the dashed lines mark their 90\% credible interval. The two-dimensional plot shows the 90\% credible regions for each waveform model, with lighter-blue shading showing the posterior distribution for the NRSur~PHM model.}
   \label{fig:remnant}
   \end{center}
\end{figure}

\subsection{Remnant Black Hole}

\paragraph{Mass and Spin} 
The merger of \ThisEvent resulted in a final (remnant) BH of mass {$M_\mathrm{f} = \mFinal\,\Msun$} (see Fig.~\ref{fig:remnant}). The inferred mass of the remnant BH provides observational evidence for an IMBH of $\gtrsim 100\,\Msun$. The remnant BH mass is {$\mTotalMinusMfinal\,\Msun$} less than the sum of the component BH masses; the equivalent energy was released as GWs 
during coalescence, making \ThisEvent the most energetic GW event recorded to date.
We find a peak luminosity close to merger 
$\ell_\mathrm{peak} = \lPeak\,\times 10^{56}\,\mathrm{erg}\,\mathrm{s}^{-1}$.

The remnant BH of \ThisEvent has a dimensionless spin parameter {$\chi_\mathrm{f} =\aFinal$}. Within the 90\% credible interval this is consistent with the BBH merger remnant spins reported in GWTC-1 \citep{LIGOScientific:2018mvr}. 
The predictions for remnant BH parameters from the inferred values of the component masses and spins agree with analyses of \ThisEvent that target the ringdown portion of the signal, directly measuring $M_\mathrm{f}$ and $\chi_\mathrm{f}$ without assuming a quasi-circular BBH, described in Section~\ref{sec:tgr}.
\paragraph{Recoil Velocity}
\begin{figure}[tb!]
\begin{center}
\includegraphics[scale=0.7]{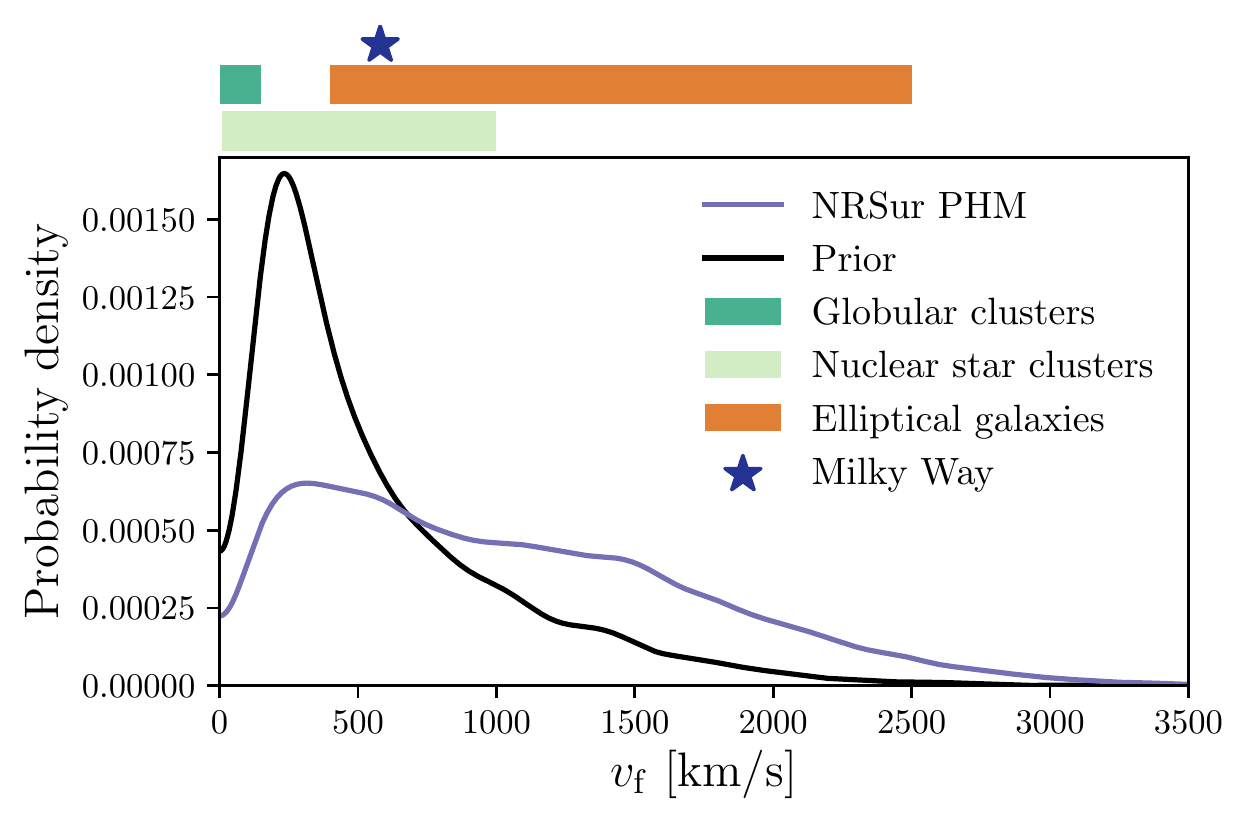}
   \caption{Posterior and prior distributions for the kick magnitude of
       \ThisEvent.  For comparison, we show known ranges for the escape
       velocities from globular clusters, nuclear star clusters, giant
       elliptical galaxies, and the Sun's location in the Milky Way.
   }
   \label{fig:kick_mag}
   \end{center}
\end{figure}

In generic BBH mergers, the radiation of linear momentum through beamed GW emission imparts a recoil
velocity, or kick, to the remnant BH \citep{1983MNRAS.203.1049F,2004ApJ...607L...5F}, of magnitude up
to ${\sim}5000$\,km\,s$^{-1}$ for precessing systems \citep{2004ApJ...607L...5F,Lousto_2011}. 
As large in-plane spin components are not ruled out for \ThisEvent, it is a potential candidate
for a large kick.  Figure~\ref{fig:kick_mag} shows the prior and posterior distributions
for the kick magnitude, with respect to the center of mass of the progenitor binary, 
computed using NRSur~PHM and applying the related remnant surrogate model \citep{PhysRevResearch.1.033015}
to the component masses and spins.  \editNew{Although the kick velocity remains unconstrained, we find the posterior
is weighted towards higher values relative to the prior,} with support for values exceeding fiducial 
escape velocities for globular clusters, the half-mass radius of nuclear star clusters 
\citep{2016ApJ...831..187A}, giant elliptical galaxies \citep{1983MNRAS.203.1049F,2004ApJ...607L...9M,2007ApJ...659L...5C}
and Milky Way-like galaxies \citep{2018A&A...616L...9M}.

\subsection{Extrinsic Parameters}

\begin{figure*}[t!]
\begin{center}
\includegraphics[width=0.90 \columnwidth]{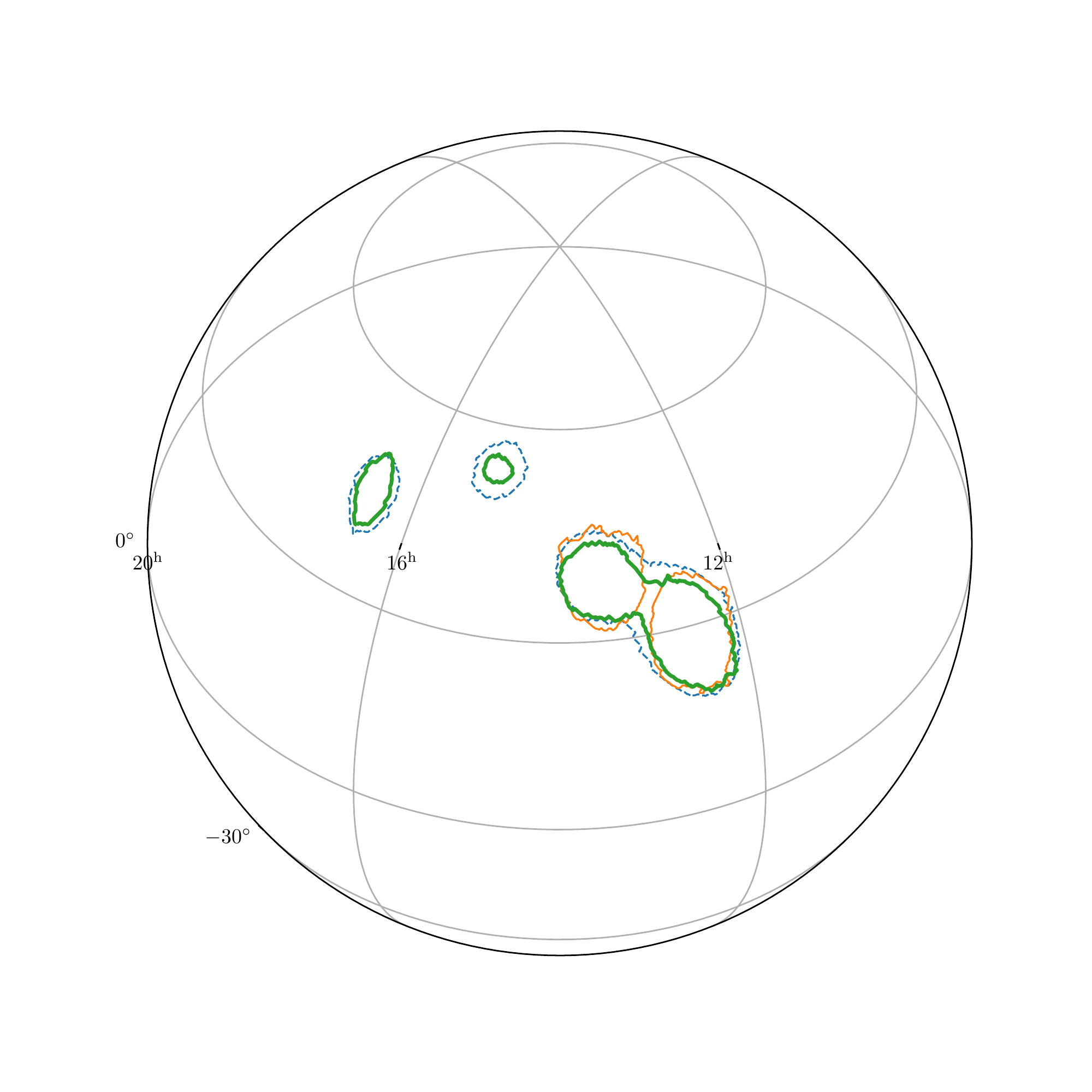}    
\includegraphics[width=0.90 \columnwidth]{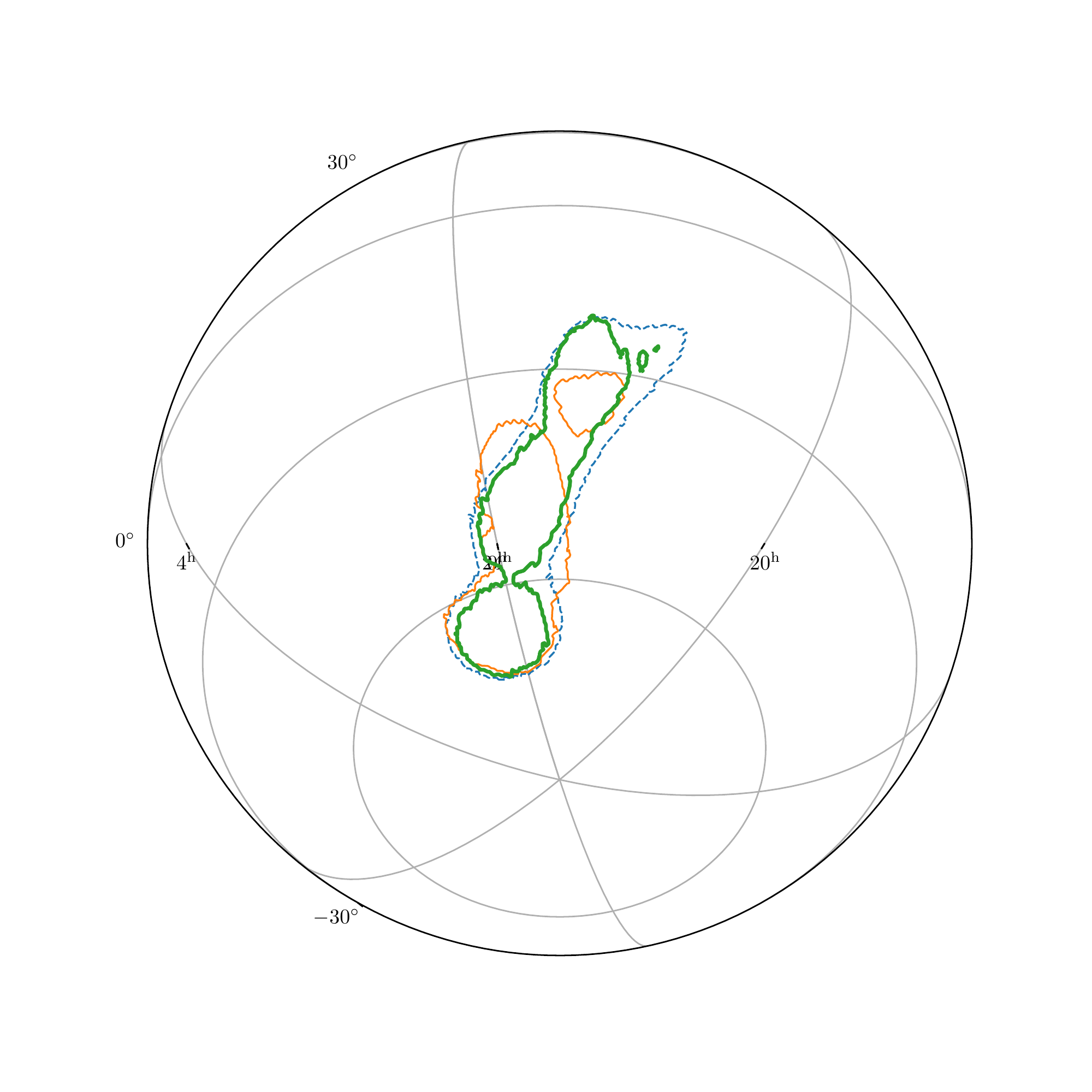}    
   \caption{Sky maps (source location 90\%\ credible areas) for \ThisEvent, as seen from the north (left) and south (right) celestial poles. 
   The blue (dashed) and the orange (solid thin) curves show two low-latency sky maps from the \texttt{BAYESTAR} pipeline and the \texttt{LALInference} pipeline using the \texttt{SEOBNRv4\_ROM} waveform model; neither incorporates higher-order multipoles. 
   The green (solid thick) curve, reported here for the first time, was obtained from full parameter estimation with the NRSur~HM model. 
   }  
   \label{fig:sky}
   \end{center}
\end{figure*}

\paragraph{Sky Position} The initial sky-map of \ThisEvent, computed using the low-latency pipeline \texttt{BAYESTAR} \citep{BAYESTAR}, was publicly released with the initial public circular on May 21, 2019 03:08:32 UTC \citep{GCN24621}. The source was localized within a 90\% credible area $\skyareaBayestar$\,deg$^2$ (see blue curve in Fig.~\ref{fig:sky}).
An updated sky-map was released the same day at 13:32:27 UTC \citep{GCN24640}, using a low-latency analysis from \texttt{LALInference} employing the \texttt{SEOBNRv4\_ROM model} \citep{PhysRevD.95.044028}, giving a 90\% credible localization within $\skyareaLalinitial$\,deg$^2$  
(orange curve in Fig.~\ref{fig:sky}). 
Here we report our latest constraints on the sky position of \ThisEvent found using \texttt{LALInference} with the NRSur~PHM model (green curve in Fig.~\ref{fig:sky}). The source is now localized within \editNew{$\skyareaLalNRSur\,$deg$^2$}. 

\begin{figure}[t!]
\begin{center}
\includegraphics[scale=0.37, trim = {0 0 0 0}]{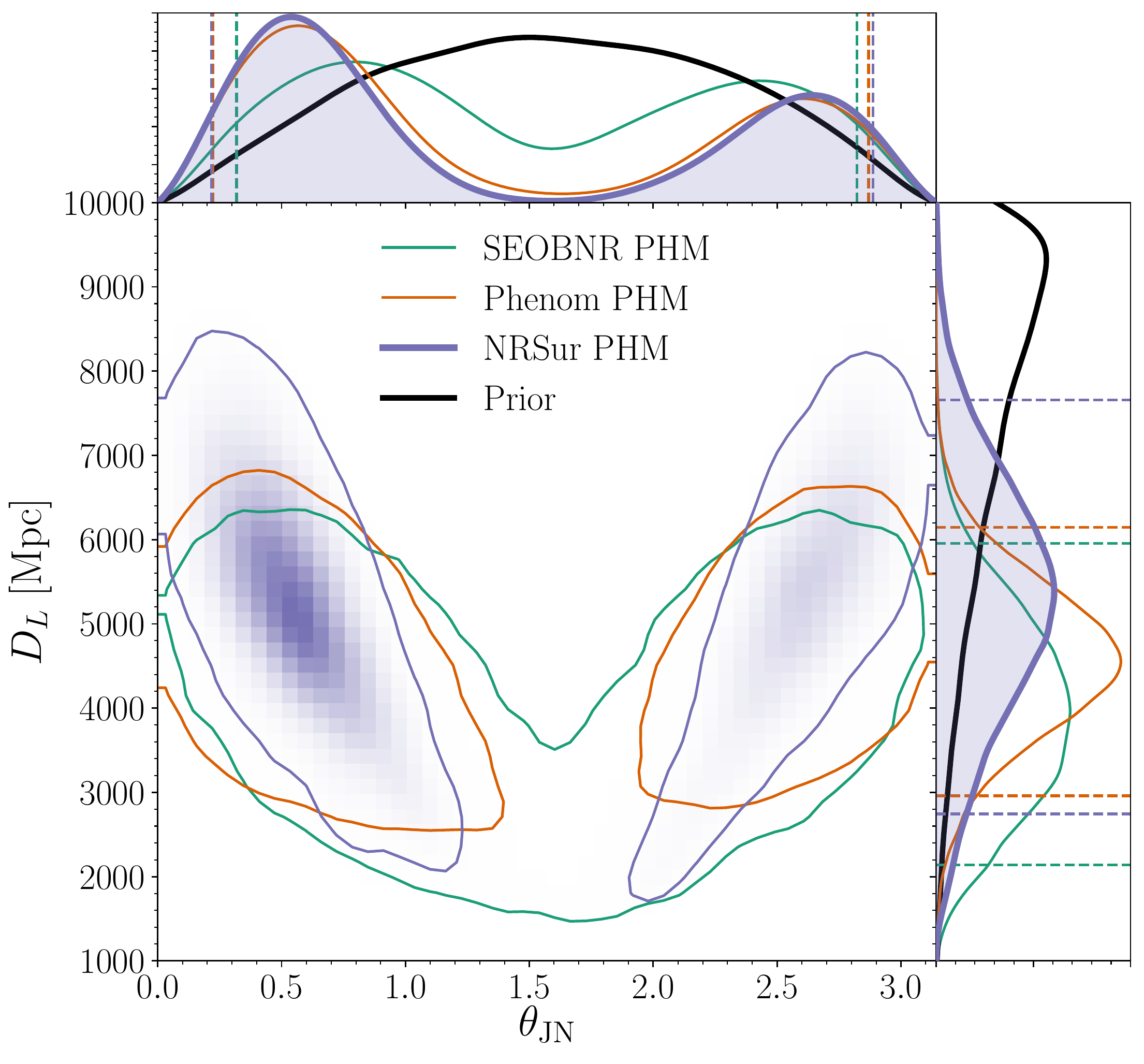}
\caption{Posterior distributions of the inclination ($\theta_\mathrm{JN}$) and the luminosity distance ($D_{L}$) according to the three waveform models employed. The one-dimensional distributions include the posteriors for the three waveform models, and the dashed lines mark their 90\% credible interval; black lines show the prior distributions. The two-dimensional plot shows the 90\% credible regions for each waveform model, with lighter-blue shading showing the posterior distribution for the NRSur~PHM model.}
   \label{fig:dist}
   \end{center}
\end{figure}

\paragraph{Distance and Inclination} 
\editNew{The inferred luminosity distance $D_L$ reported at low latency \citep{GCN24621,GCN24640} made use of waveform models that did not include higher-order multipoles. As discussed in Section 2.5, the luminosity distance and redshift inferred using the waveform models described in this paper which incorporate higher-order multipoles result in larger values than those low-latency estimates.}
In Figure~\ref{fig:dist}, we show the posteriors on the luminosity distance $D_L$ and the inclination angle $\theta_\mathrm{JN}$ of \ThisEvent for the three waveform models. Here, the inclination angle is between the {\it total} angular momentum of the source, and the observer's line-of-sight. One can note in the figure the typical correlation between $D_L$ and $\theta_\mathrm{JN}$, and near degeneracy between systems with the angular momentum vector pointed towards or away from the observer. 
GW emission is strongest along the orbital \editNew{angular} momentum direction, so face-on sources at larger distances produce similar signals to edge-on sources closer by.
Both the NRSur~PHM and Phenom~PHM models suggest that the total angular momentum of the source is roughly aligned with the observer's line of sight $\theta_\mathrm{JN} \sim 0$ or $180^\circ$, i.e.\ the orbital plane is close to face-on to the line of sight, while the SEOBNR~PHM model also supports an orbital plane that is closer to edge-on ($\theta_\mathrm{JN} \sim 90^\circ$).  In accordance with the covariance between the distance and inclination, the NRSur~PHM model places \ThisEvent at \editNew{$D_L=\LuminosityDistance$\,Gpc ($z \simeq \redshift$)} while the SEOBNR~PHM model suggests \editNew{$D_L=\LuminosityDistanceSEOBNR$\,Gpc}. 
The masses of the component and remnant BH scale inversely with $(1+z)$. In accordance, as reported in Table~\ref{tab:parameters}, the SEOBNR~PHM model supports higher masses than the other two models.
A close to face-on binary, as favored by our preferred NRSur~PHM model, makes it difficult to directly observe the effects of spin-orbit precession; thus, the evidence for precession in \ThisEvent, reported in the {\it Spins} subsection of Section~\ref{sec:PE_components} above, is weak.

\subsection{Impact of Higher-Order Multipoles}

\begin{figure}[tb!]
\begin{center}
\includegraphics[scale=0.37, trim = {0 0 0 0}]{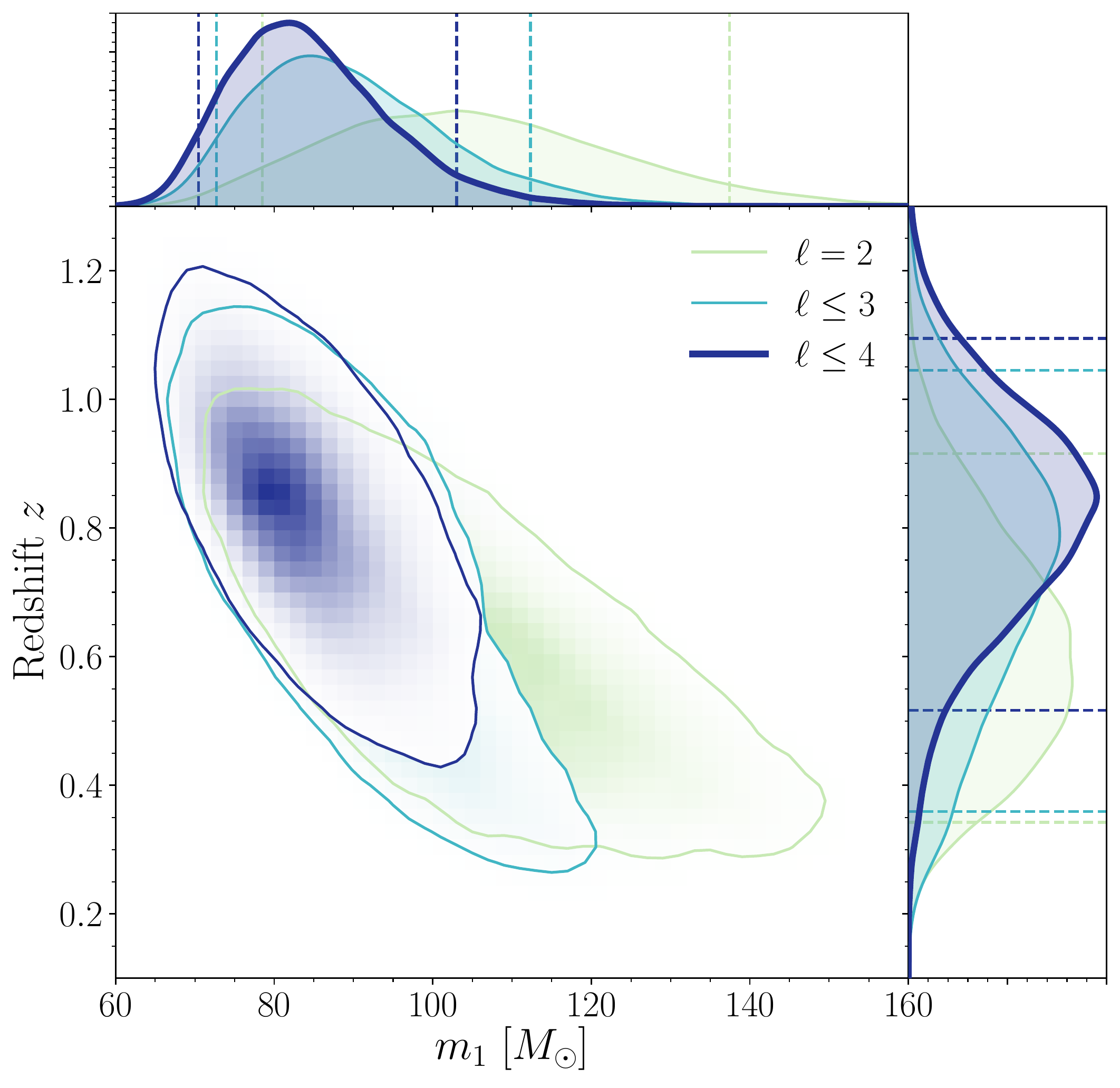}
   \caption{Posterior probability density for the source-frame mass of the primary BH $m_1$ and luminosity distance $D_L$. The one-dimensional distributions include the posteriors for the NRSur~PHM waveform using the \texttt{RIFT} pipeline, for different combinations of multipoles: dominant multipole $\ell=2$ (light green), higher-order multipoles up to $\ell\leq3$ (light blue) and those up to $\ell\leq4$ (dark blue). The two-dimensional plot shows the contours of the 90\% credible areas, with the light-blue shading showing the posterior density function for $\ell\leq4$.}

   \label{fig:HM}
   \end{center}
\end{figure}
The GWs emitted during a BBH coalescence can be decomposed as spherical harmonic multipoles $\ell$ and $|m| \leq \ell $. The quadrupole 
$(\ell, |m|) = (2, 2)$ mode remains dominant during most of the inspiral, while other, higher-order multipoles become comparable near the merger and ringdown stages (see for instance \citealt{Bustillo:2015qty,SEOBNR.Multipolar}). 
Higher-order multipoles have been found to be important for high mass BBH mergers in advanced detector observations \citep{Varma:2014jxa,Varma:2016dnf,Graff_2015,Bustillo:2016gid, CalderonBustillo:2017skv,2019PhRvD.100j4015C,SEOBNR.Multipolar,Mehta:2017jpq,Mehta:2019wxm}; their presence was also crucial in reducing uncertainties in the mass ratio and component spins of the recently reported unequal-mass binary mergers GW190412 \citep{LIGOScientific:2020stg} and GW190814 \citep{abbott190814}. 
To quantify the impact and evidence of higher-order multipoles on \ThisEvent, we computed the posterior distribution of our preferred model, NRSur~PHM \citep{PhysRevResearch.1.033015}, using the \texttt{RIFT} pipeline \citep{Lange:2018pyp} for three different combinations of modes: $\ell=2$, $\ell\leq 3$ and $\ell\leq 4$.  

 \editNew{GR predicts radiation from BBH mergers at all multipoles, with amplitudes strongly dependent on the binary mass ratio and orbital inclination angle \citep{Graff_2015,London:2017bcn}; 
evidence of their presence consistent with GR is presented in \citet{LIGOScientific:2020stg}.}

We find that the omission of higher-order multipoles leads to broader posterior distributions for some parameters of \ThisEvent, especially the binary orbital inclination angle $\theta_\mathrm{JN}$. 
The higher-order multipoles enable better constraints of the binary inclination angle, which is coupled to the source-frame mass estimates through the luminosity distance $D_\mathrm{L}$, and therefore the redshift. 
\editNew{Inclusion of these higher-order multipoles results in significantly larger values for these parameters.}
As shown in Fig.~\ref{fig:HM}, this change in distance directly impacts our inference of the primary BH mass of \ThisEvent. 
For our preferred model, if we only included $\ell=2$, the posterior distribution of primary BH mass extends to {$m_1 = \mOneNoHOM\,\Msun$} at 90\% credible interval. As we include higher-order multipoles up to $\ell=4$, the masses are better constrained and we find that $m_1$ cannot be greater than {$\mOneHOM\,\Msun$} at 90\% credible interval. 
We find that the higher-order multipoles have marginal impact on constraining the effective precession spin parameter $\chi_\mathrm{p}$.

 \editNew{Although the released version of NRSur PHM \citep{PhysRevResearch.1.033015} does not include multipoles with $\ell > 4$, we have extended it to include all multipoles with $\ell = 5$, and find that the parameter estimation results are nearly identical to those for $\ell \le 4$; the $\ell = 5$ multipoles are found to be quantitatively negligible for \ThisEvent.}

While higher-order multipoles influence our inference of the source properties, the observation of \ThisEvent does not offer evidence for such multipoles. The $\log_{10}$-Bayes factor for our preferred model with and without higher-order multipoles is $\logTenBFHOM$, \editNew{implying that the data marginally disfavors their presence, but there is no statistically significant evidence for their absence}.  \editNew{As noted above, this gives important information about the orbital inclination of the binary, as higher-order multipoles are suppressed for face-on binary orbits.}
Since higher-order multipoles are a firm prediction of GR and significantly affect parameter estimation, they must be included when modeling \ThisEvent.

\subsection{Comparison with NR}

We compared the data from LIGO and Virgo at the time of \ThisEvent directly with {3459} NR simulations of BBH coalescence \citep{Jani_2016, Healy_2019, Boyle_2019}. These NR simulations are the most accurate representation available of the strong-field dynamics near merger and the radiated higher-order multipoles. 
As was done for previous direct comparisons of NR simulations with LIGO-Virgo detected events \citep{PhysRevD.94.064035,Healy_2018, LIGOScientific:2018mvr}, the likelihood of the data was calculated for each waveform derived from NR simulations.  Using the \texttt{RIFT} pipeline \citep{Lange:2018pyp} to evaluate these comparisons and interpolate them over all intrinsic parameters, we deduce a posterior distribution for all detector-frame quantities: redshifted mass, mass ratio, and both component spins. 

\begin{figure}[tb!]
\begin{center}
\includegraphics[scale=0.37, trim = {0  0 0 0}]{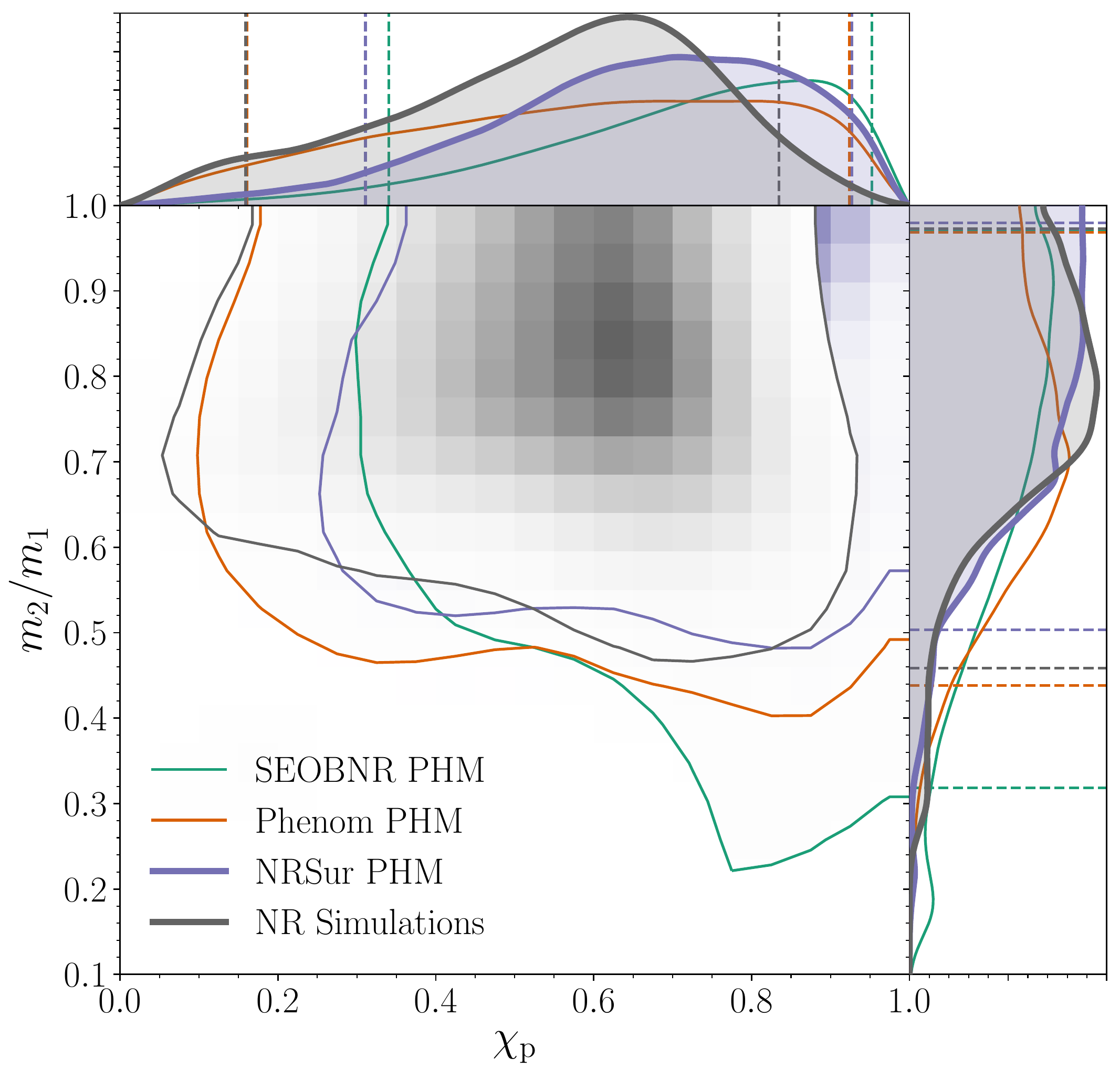}
   \caption{Posterior probability density for the precessing spin parameter $\chi_\mathrm{p}$ and mass ratio from the direct comparison of NR simulations with \ThisEvent (gray tiles, and gray contour showing 90\% confidence area). For comparison, shown here are the 90\% bounds on these parameters from the three waveform models.} 
   \label{fig:NR}
   \end{center}
\end{figure}

As shown in Fig.~\ref{fig:NR}, we find that the best matching (highest likelihood) NR waveforms and the derived posterior distribution are consistent with the mass-ratio of \ThisEvent being near unity.  We also find the analysis favors NR waveforms with
$\chi_\mathrm{eff} \sim 0$ and $\chi_\mathrm{p}\sim 0.6$, further suggesting that \ThisEvent is a precessing binary. The agreement between our preferred model NRSur~PHM and the NR simulations provides an independent check on the inferred source properties of \ThisEvent. 

\section{Consistency with Binary Merger Waveform Models}

\ThisEvent{} presents an opportunity for strong-field tests of general relativity in a previously unexplored region of parameter space \citep{TGR_paper_GW150914, LIGOScientific:2019fpa}.  We first test the consistency of the data with the waveform models employed for parameter estimation by searching for unmodelled residual power; we then study in detail the properties of the ringdown phase using fewer assumptions on the progenitors of the remnant BH, and demonstrate consistency with predictions from the full waveform models.

\subsection{Residual Tests}

We first test the consistency of the templates used for parameter estimation with the observed signal by subtracting the maximum likelihood NRSur~PHM waveform from the data and studying the residual \citep{TGR_paper_GW150914,LIGOScientific:2019hgc}. 
The residual data is analyzed by the {\tt BayesWave} algorithm which simultaneously fits models for the noise PSD and short ($<1\,$s) transient signals coherent across the detector network \citep{Cornish:2014kda}. 
{\tt BayesWave} does not assume a particular waveform morphology, instead using a linear combination of wavelets to fit excess coherent features in the data of arbitrary shape.

The metric used for quantifying the coherent residual, $\mathrm{SNR}_{90}$, is the upper 90\% credible bound on the posterior distribution function of the recovered signal-to-noise ratio of the wavelet reconstruction \citep{TGR_paper_GW150914}.
The significance of the recovered $\mathrm{SNR}_{90}$ value is empirically measured by repeating the analysis on $\BayesWaveNumOffsource$ randomly chosen off-source times drawn from 
$\BayesWaveOffsourceTime$ seconds of detector data surrounding the event time, which serves as the background estimate for $\mathrm{SNR}_{90}$. 
For \ThisEvent, we find $\mathrm{SNR}_{90}\sim \BayesWaveResidualSnrNinety$, which is consistent with expectations for 
typical LIGO-Virgo noise, resulting in a p-value of ${\sim} \BayesWaveResidualPvalue$ when compared to the times analyzed immediately surrounding the merger time. 
Since the residual is consistent with noise, we find that the best fit waveform interpolated from numerical solutions of general relativity, as used in our parameter estimation analyses, is consistent with the observed signal within the measurement capability of the detectors.

\subsection{Tests using final black hole ringdown}
\label{sec:tgr} 

In this section, we describe a number of studies that explore the nature of the remnant compact object and also evaluate the consistency of the observed signal with waveform models for a quasi-circular BBH merger in GR. Thse are extensions to similar tests applied to previous LIGO-Virgo BBH events \citep{TGR_paper_GW150914, LIGOScientific:2019fpa}. As in the case for those previous tests, the high mass of \ThisEvent\ makes it possible to study the quasi-normal oscillations of the remnant BH approaching a stationary state (ringdown), as encoded in the last few cycles of the GW signal; see \citet{Vishveshwara, Buonanno:2006ui, Berti:2009kk, Berti:2018vdi} and references therein. Parameter estimates obtained from the ringdown studies presented in this section are solely extracted from the properties of the remnant. 
For these studies, we consider both a model which makes no assumptions on the process leading to the signal formation, and also waveform templates specifically modelling the remnant of a quasi-circular BBH merger. The results are thus robust against systematic uncertainty due to the possible neglect of eccentricity or other physical effects in modelling the system; for further discussion of such effects see Section~\ref{sec:alt_ecc}.

We model the ringdown as a set of damped sinusoids and measure the properties of the signal using {\tt pyRing}, a time domain analysis framework \citep{Carullo:2019flw, 2019arXiv190500869I}.
Within the analysis, the beginning of the ringdown-dominated portion of the signal is marked by a fixed time $t_0$ (see Supplement of \citealt{2019arXiv190500869I}), reported with respect to an estimate of the peak time of the complex strain at each detector. The peak time is taken from the posterior median inferred from the NRSur~PHM model and yields a GPS time for LIGO Hanford $t^{\rm H}_{\rm peak} = \PeaktimeH \,\mathrm{s}$; times in other detectors are computed by assuming a fixed sky position. The duration of the analysis segment after the start time in each detector is $0.1$\,s.  
We consider three distinct models of the ringdown: for each, we use the {\tt CPNest} Bayesian nested sampling algorithm \citep{CPNest} to compute posterior distributions on the model parameters.

\begin{figure*}[tb!]
\centering
\includegraphics[width=0.45\textwidth]{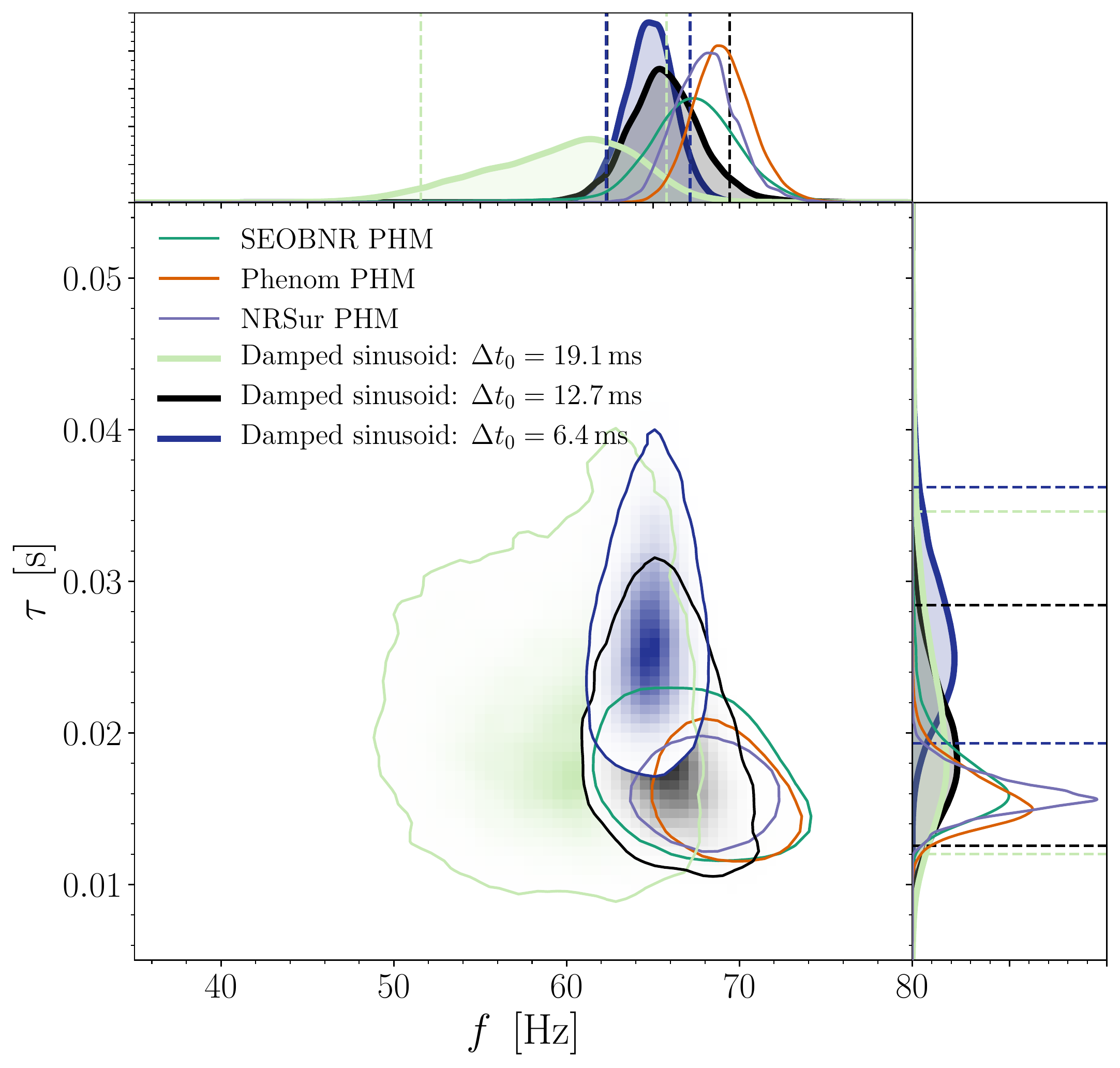}
\includegraphics[width=0.45\textwidth]{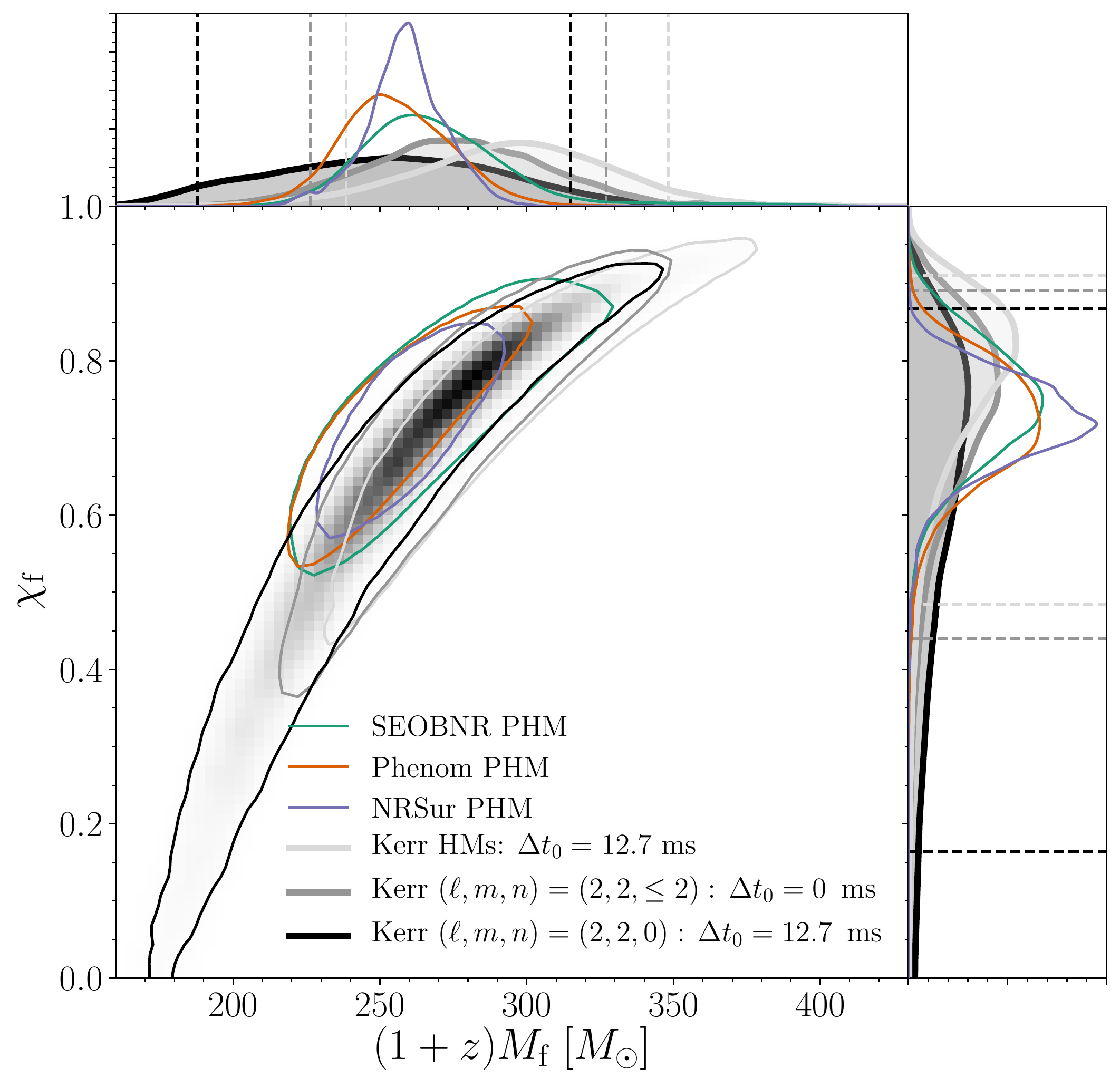}
    \caption{Left: redshifted (detector-frame) frequency and damping time inferred from the ringdown portion of the \ThisEvent signal. 
    Measurements using a single damped sinusoidal model of the ringdown are shown with filled contours at different start times $\Delta t_0 = \PeaktimeFiveM$\,ms (blue), $\PeaktimeTenM$\,ms (black), and $\PeaktimeFifteenM$\,ms (light-green) (${\sim} 5, 10, 15 \,{}GM^\mathrm{det}_\mathrm{f}/c^3$) after the reference $t^{\rm H}_{\rm peak}$. These are compared with the least-damped ringdown mode from the three distinct inspiral-merger-ringdown waveform models described in Section \ref{sec:PE-method}.
    Right: redshifted remnant mass $M^\mathrm{det}_\mathrm{f} = (1 + z) M_\mathrm{f}$ and spin inferred from the ringdown portion of the signal. 
    The filled contours show the measurement using the fundamental Kerr $\ell=2,m=2,n=0$ multipole (black); the 
    $\ell=2,m=2$ Kerr model including overtones up to $n=2$ (gray); and the fundamental higher mode model (Kerr HMs, light gray) described in the text. 
    These are compared with redshifted remnant mass and spin obtained using the three waveform models stated in Section \ref{sec:PE-method} and Fig.~\ref{fig:remnant}.
    Contours enclose 90\% of the posterior distribution, and the 1D histogram shows the 90\% credible regions only for the ringdown models.
    }
\label{fig:tgr_rd}
\end{figure*}
The first model consists of a single damped sinusoid of arbitrary complex frequency, amplitude and phase, with no assumptions on the nature of the remnant object.
The left panel of Fig.~\ref{fig:tgr_rd} shows the resulting 90\% credible regions for the 
posterior probability distribution of the redshifted, detector frame frequency and damping time, assuming uniform priors on these two parameters (solid curves). 
Since there is uncertainty in the time at which the single damped sinusoid model may become valid, we perform this analysis by truncating the data at different start times, as in \cite{Kamaretsos:2011um} and \cite{TGR_paper_GW150914}: we choose start times $\Delta t_0 \equiv t_0 - t^{\rm H}_{\rm peak} \simeq \PeaktimeFiveM , \PeaktimeTenM, \PeaktimeFifteenM$\,ms after the reference time.
In units of the redshifted remnant mass $M^\mathrm{det}_\mathrm{f} = (1 + z) M_\mathrm{f} = \FinalMassNRSurMedian \,\Msun$ (median estimate from NRSur~PHM) these times correspond to $\Delta t_0 = 5, 10, 15 \,GM^\mathrm{det}_\mathrm{f}/c^3$. Truncating the data at later times yields uninformative posteriors due to the decrease in SNR. Values of $\Delta t_0 \geq 10\,GM^\mathrm{det}_\mathrm{f}/c^3$ are consistent with the range where numerical simulations predict that the fundamental mode should be dominant for this source. 
The systematic uncertainty due to the choice of the NRSur~PHM model in determining $t_{\rm peak}$ as opposed to other waveform models considered in this paper is negligible with respect to the statistical uncertainty. 

At a start time of $\Delta t_0 = \PeaktimeTenM$\,ms (corresponding to $10\,GM^\mathrm{det}_\mathrm{f}/c^3$), we measure a ringdown frequency $f=\FreqRing$\,Hz and damping time $\tau= \TauRing$\,ms.
As the start time increases from $\PeaktimeFiveM$ to $\PeaktimeFifteenM$\,ms, we observe convergence towards the predicted value of the least-damped mode frequency, together with a broadening of the posterior due to the decreasing SNR of the ringdown. Similar behavior is observed for the other models we will present in the remainder of the section. 

The second model is built from the superposition of a set of damped sinusoids with arbitrary amplitudes and phases, but with complex frequencies determined by the remnant mass $M_\mathrm{f}$ and dimensionless spin $\chi_\mathrm{f}$, as predicted by perturbation theory \citep{Berti:2005ys}.
We include up to the $n=0,1,2$ overtones of the $\ell=m=2$ ringdown mode of a Kerr BH starting at the peak of the complex strain \citep{Buonanno:2006ui,Giesler:2019uxc,Ota:2019bzl}; see also \citet{Bhagwat:2019dtm,Okounkova:2020vwu} for discussions concerning the interpretation of a linearised approximation starting at the peak of GW emission.
The redshifted, detector frame remnant mass and spin obtained from this waveform model are shown in the right panel of Fig.~\ref{fig:tgr_rd}, assuming a uniform prior over these two parameters. We find $M^\mathrm{det}_\mathrm{f}=\FinalMassRing \,\Msun$ and $\chi_\mathrm{f}= \FinalSpinRing $ taking $\Delta t_0 = \PeaktimeTenM$\,ms and including only the fundamental Kerr $(\ell=2,m=2,n=0)$ mode; and $M^\mathrm{det}_\mathrm{f} = \FinalMassTwoOvertones \,\Msun$, $\chi_\mathrm{f}=\FinalSpinTwoOvertones$ taking $\Delta t_0 = 0$\,ms and including overtones up to $n=2$.

Finally, the third model (Kerr HMs) consists of a set of damped sinusoids corresponding to all fundamental modes (i.e.\ without inclusion of overtones) of a Kerr BH up to $\ell=4$ and $m=\ell,\,\ell-1$, including spherical-spheroidal harmonic mixing \citep{London:2018gaq}. Complex frequencies are predicted as a function of the remnant mass $M_\mathrm{f}$ and dimensionless spin $\chi_\mathrm{f}$, while amplitudes and phases are calibrated on numerical relativity simulations of non-precessing BBH mergers.
With this model we measure $M^\mathrm{det}_\mathrm{f}=\FinalMassHM\,\Msun$, $\chi_\mathrm{f}=\FinalSpinHM$ taking $\Delta t_0 = \PeaktimeTenM$\,ms; the full probability distribution is shown in the right panel of Fig.~\ref{fig:tgr_rd}.

In Figure~\ref{fig:tgr_rd}, we compare the ringdown measurements to the \editNew{posterior credible regions for the remnant parameters} obtained through NR-calibrated fits from the initial binary parameters, as described in Section~\ref{sec:PE-method}, from the three different full waveform models discussed above.
The posterior of the ringdown analyses is consistent at the $90\%$ credible level with the full-signal analyses.
Furthermore, despite the different physical content, both models that include higher multipoles or overtones obtain measurements of remnant parameters consistent with the single-mode analysis estimates; a Bayes factor computation also does not find strong evidence in favour of the presence of higher multipoles or overtones.

A parametrized test of gravitational waveform generation \citep{2012PhRvD..85h2003L,2012JPhCS.363a2028L,2014PhRvD..89h2001A,Blanchet:1994ez,Mishra:2010tp} based on the Phenom~PHM waveform model, also does not reveal inconsistencies with GR predictions. Full details will be provided in an upcoming paper.

\section{Single-Event Based Merger Rate Estimate}
We estimate the rate of mergers similar to this source, assuming a constant rate per comoving volume-time element.  We proceed similarly to 
\cite{GW150914:rates}: in the absence of a parameterized population model for such sources, we assume a population of mergers whose intrinsic 
parameters (component masses and spins) are identical to the detected event up to measurement errors \citep{Kim2003}.  We estimate the 
sensitivity of the LIGO-Virgo detector network by adding simulated signals to data from the O1, O2 and O3a observing runs and recovering 
them with the \texttt{Coherent WaveBurst} (cWB) weakly-modeled transient detection pipeline \citep{PhysRevD.93.042004}, optimized for sensitivity to IMBH mergers 
\citep{Virgo:2012aa}, which identified \ThisEvent with the highest significance \citep{GW190521-Discovery}.  As in \cite{Abbott:2018exr} and \cite{O1BNS}, we consider a simulated signal to be detected if recovered
with an estimated false alarm rate of 1 per 100 years or less.  
Simulated signals are prepared by drawing source parameter samples from a posterior distribution inferred using the NRSur~PHM waveform (Section~\ref{sec:PE}).  The simulations' 
component masses and spins are taken directly from the posterior samples, whereas their line-of-sight direction and orbital axis direction are randomized, and their luminosity distances 
are distributed uniformly over comoving volume and time. 

The time- and angle-averaged sensitive luminosity distances of the cWB-IMBHB search for mergers similar to \ThisEvent in O1, O2 and O3a data are {\cWBOoneDistance\,Gpc}, {\cWBOtwoDistance\,Gpc} and {\cWBOthreeaDistance\,Gpc}, respectively, showing a substantial gain in sensitivity over successive runs. 
The combined searched time-volume over O1, O2 and O3a data is {$\VT\,\mathrm{Gpc}^3\,{}\mathrm{yr}$}.  Thus, taking a Jeffreys prior on the rate $R$ as $p(R)\propto R^{-0.5}$, and with 1 event detected above threshold, we obtain an 
estimate {$R = \RateMedian^{+\RatePlus}_{-\RateMinus}\,\mathrm{Gpc}^{-3}\,{}\mathrm{yr}^{-1}$}.  

This rate is below previous upper limits obtained by various methods: (mass- and spin-dependent) upper limits inferred from LIGO-Virgo searches of O1-O2 data, $0.20\,$Gpc$^{-3}$\,yr$^{-1}$ or greater \citep{2019PhRvD.100f4064A}; limits obtained in \citet{ch2020nuria} from numerical simulations of BBH signals with generic precessing spins added to O1 data, $0.36$\,Gpc$^{-3}$\,yr$^{-1}$ or greater.  Our rate is also well below model-dependent limits of $\mathcal{O}(1)\,$Gpc$^{-3}$\,yr$^{-1}$ obtained in \citet{2019arXiv191105882F} for possible BBH populations with $45 < m_1/\Msun < 150$. 
Our estimate is for a population sharing the properties of \ThisEvent; as we obtain more observations of high mass BBH systems we expect to better constrain the distribution of their masses and spins, and thus refine the population rate estimate. 
The merger rate obtained here may be compared to expectations from possible formation channels, as in the following section.

\section{Astrophysical Formation Channels and Implications for Stellar Collapse}

Both the primary component mass and remnant mass of this system are higher than the previously most massive BBH detected by LSC and Virgo, GW170729 \citep{LIGOScientific:2018mvr}.  Other candidate events, dubbed GW170817A and GW151205, have been identified in O1 and O2 data by other groups \citep{Zackay:2019btq,Nitz:2019hdf}, with component masses higher than GW170729 if of astrophysical origin.  Here we do not attempt to assess possible astrophysical implications of such additional events.

Our analysis of the BBH population detected in the O1 and O2 runs using parameterized models of the mass distribution indicates that $99\%$ of merging BH primaries have masses below $45\,\Msun$ \citep{2019ApJ...882L..24A,2017ApJ...851L..25F}.  As the primary mass of \ThisEvent is well above this value, the system is well within the highest mass $1\%$ of the population inferred from O1 and O2 observations. We obtain confirmation of the unexpected nature of \ThisEvent by generating synthetic catalogs of 25 (50) BBH detections to represent expectations for BBH detections in the first half of O3. The masses of synthetic detections are obtained via draws from the posterior of the O1--O2 population described by the most general model, Model C, of \cite{2019ApJ...882L..24A}, after applying selection effects. We then extract the highest-mass primary BH $m_{1,\mathrm{max}}$ and compare to the primary mass of \ThisEvent. For the 25-event (50-event) synthetic catalogs, \twentyFiveEventsProbGtMassOneMedian\ (\fiftyEventsProbGtMassOneMedian) of $m_{1,\mathrm{max}}$ values lie above \mOneMedian\ and \twentyFiveEventsProbGtMassOneFivepct\ (\fiftyEventsProbGtMassOneFivepct) lie above \mOneFivePct\ (the posterior median and 5th percentile estimated using the NRSur~PHM model). 
Even when accounting for statistical uncertainties in the mass estimation \citep{2019arXiv191105882F}, the primary mass of \ThisEvent is in tension with the population inferred from O1 and O2.

\begin{figure*}[t!]
\begin{center}
\includegraphics[width=1.6\columnwidth]{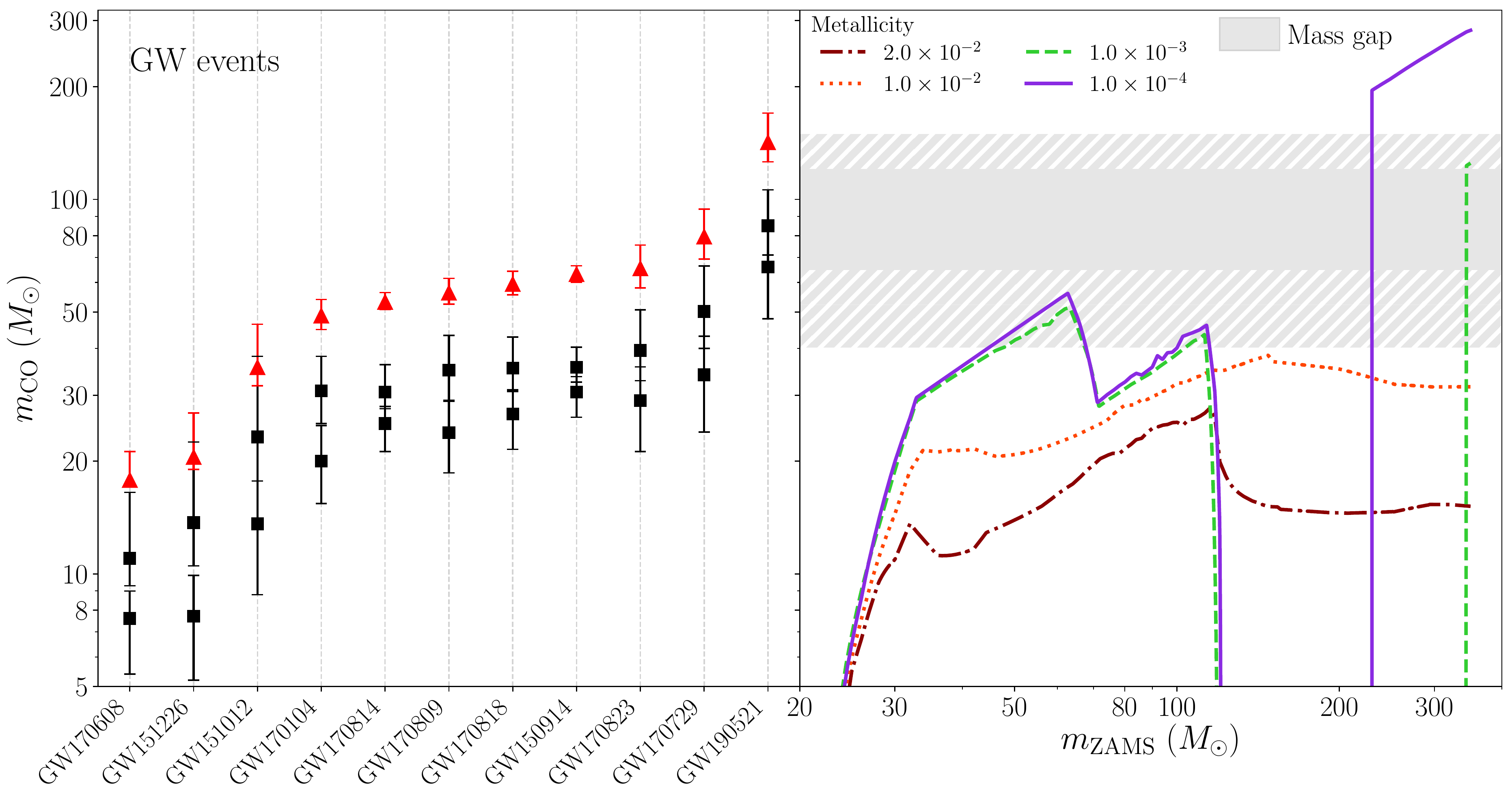}
\caption{Left: Masses ($m_{\rm CO}$) of \ThisEvent compared with BBH detections in O1 and O2. Black squares and error bars represent the component masses of the merging BHs and their $90\%$ uncertainties, and red triangles represent the mass and associated uncertainties of the final merger remnants. Right: Predicted compact-object mass, as a function of the zero-age main sequence mass of the progenitor star ($m_{\rm ZAMS}$ up to 350 M$_\odot$) and for four different metallicities of the progenitor star (ranging from $Z=10^{-4}$ to $Z=2\times 10^{-2}$, \citealt{2017MNRAS.470.4739S}). This model accounts for single stellar evolution from the PARSEC stellar-evolution code \citep{2012MNRAS.427..127B}, for core-collapse supernovae \citep{Fryer:2011cx}, and for PPI and PI supernovae \citep{2017ApJ...836..244W}. Only the two metal-poor models with metallicity $Z=10^{-3}$ (dashed green line) and $Z=10^{-4}$ (purple solid line) undergo PI supernovae and leave no compact objects between $m_{\rm ZAMS}\sim{}119-344$  M$_\odot$ and $\sim119-230$ M$_\odot$, respectively.  The shaded area shows the PI mass gap, the hatched regions corresponding to the uncertainty in current models (e.g.,\  \citealt{2019ApJ...887...53F,2020ApJ...888...76M}).}
   \label{fig:figPISN}
   \end{center}
\end{figure*}

Figure~\ref{fig:figPISN} shows the mass of \ThisEvent in comparison with the masses of all the O1 and O2 BBHs (left-hand panel) and with current theoretical knowledge about the PI mass gap (right-hand panel). A set of evolutionary astrophysical models \citep{2017MNRAS.470.4739S} relating the progenitor mass and compact object is also shown for reference. The astrophysical models are subject to several uncertainties in stellar evolution and core-collapse supernovae, and only serve as representative examples.  This figure is an update of Fig.~5 of \cite{2019ApJ...882L..24A} to include the masses of GW150521.1 and the most recent uncertainty estimates on the PI mass gap, which we review below in Section~\ref{sec:PI_uncertainty}. 
%

Several mechanisms could fill the PI gap. In the following sections, we discuss second-generation (2g) BHs, stellar mergers in young star clusters and BH mergers 
in AGN disks.  Second-generation BHs, i.e.\ BHs born from the merger of two BHs, can have mass in the PI gap \citep{2002MNRAS.330..232C,2017PhRvD..95l4046G,2017ApJ...840L..24F,2019PhRvD.100d3027R}. If they are in a dense stellar environment and are not ejected by the gravitational recoil, they have a chance to form a new binary system with another BH \citep{2019PhRvD.100d3027R,2019PhRvD.100d1301G}. Alternatively, a merger between an evolved star (with a well developed helium or carbon-oxygen core) and a main-sequence companion might trigger the formation of a giant star with an over-sized envelope with respect to the core. If this star collapses into a BH before its helium core enters the PI range, then it can give birth to a BH in the PI gap  \citep{2019MNRAS.485..889S,2019MNRAS.487.2947D}. 
If this BH is inside a dense stellar environment, it has a chance to capture a companion by dynamical exchange. BHs in AGN disks might pair with other BHs and are expected to merge efficiently due to gas torques, producing 2g BHs (e.g.,\ \citealt{2018ApJ...866...66M}). These BHs in AGN disks might even grow by gas accretion (e.g.,\ \citealt{2012MNRAS.425..460M}).

Finally, \cite{2020arXiv200405187V} investigated the possibility that BHs in isolated binaries might have mass in the PI gap by super-Eddington accretion of the companion onto the primary BH. Even under the most extreme assumptions, they find that at most $\sim{}2$\% of all merging BBHs in isolated binaries contains a BH with a mass in the pair-instability mass gap, and find no merging BBH with a total mass exceeding $100\,\Msun$. 

\subsection{Uncertainties on PI and stellar evolution theory}
\label{sec:PI_uncertainty}
The physical processes leading to PI and PPI are well known, and a robust formalism to model them has been developed (e.g.,\ \citealt{1964ApJS....9..201F,1967PhRvL..18..379B, 1983A&A...119...61O,1984ApJ...280..825B, 2003ApJ...591..288H,2007Natur.450..390W,2014ApJ...792...28C,2016MNRAS.457..351Y,2017ApJ...836..244W,2019ApJ...878...49W,2019ApJ...882...36M}). 
On the other hand, there are still uncertainties on the minimum helium-core mass for a star to undergo PPI: \cite{2017ApJ...836..244W} indicate $M_{\rm He,\,{}min} \approx 32\,\Msun$, while \cite{2019ApJ...887...72L} suggest that this value should be raised to $M_{\rm He,\,{}min}\approx 40\,\Msun$. This difference might translate into a significant difference in the maximum BH mass. 

In addition to this, there are critical uncertainties on other physical ingredients which combine with PI and PPI to shape the mass spectrum of BHs (e.g.,\ \citealt{2016A&A...594A..97B,2019ApJ...882..121S,2020A&A...636A.104B}). Recently, \cite{2019ApJ...887...53F} and \cite{2020MNRAS.tmp..537R} have investigated the main sources of uncertainty on the lower edge of the PI mass gap, by modelling naked helium cores. The impact of time dependent convection on the location of the lower edge of the PI mass gap is found to be $\Delta m\approx 5\,\Msun$ \citep{2020MNRAS.tmp..537R}. Variations in the treatment of convective mixing and  neutrino physics are found to have small effects on the maximum BH mass (with a mass variation $\Delta m\approx 1-2\,\Msun$), while different stellar metallicity and wind mass loss prescriptions have more conspicuous repercussions ($\Delta m\approx 4\,\Msun$). Most importantly, the  uncertainties on nuclear reaction rates have a dramatic impact on the maximum BH mass, which can change  by $\Delta m\approx 16\,\Msun$, corresponding to $30-40\%$ of the maximum BH mass. Most of the differences come from 
the $^{12}$C$(\alpha,\,\gamma)\,^{16}$O reaction (\citealt{2019ApJ...887...53F}; see also \citealt{2018ApJ...857..111T}). This result is particularly important if we consider that \cite{2019ApJ...887...53F} explore only $1\,\sigma$ uncertainties \editNew{(see, e.g., \citealt{2017RvMP...89c5007D} for a recent review)} and vary only a few nuclear reaction rates. 

The maximum BH mass estimated by \cite{2019ApJ...887...53F} is ${\approx}56\,\Msun$; however they model pure helium cores, without hydrogen envelopes. \cite{2020ApJ...888...76M} show that accounting for the collapse of a residual hydrogen envelope can increase the maximum BH mass by ${\approx}20\,\Msun$ in the case of metal-poor ($Z\leq{}0.0003$), slowly rotating progenitor stars. Hence, the final fate of the hydrogen envelope is an additional source of uncertainty, with an impact of $\Delta m\approx 20\,\Msun$  on the maximum BH mass.

While mass transfer in close binaries is expected to remove most of the hydrogen envelope, metal-poor ($Z\leq 3\times 10^{-4}$)  single stars with zero-age main sequence (ZAMS) mass $m_{\rm ZAMS}\lesssim 70\,\Msun$ are expected to retain a significant fraction of their hydrogen envelope. If most of this hydrogen envelope collapses into the final compact object, the maximum BH mass is ${\sim}60-65\,\Msun$, while a naked helium core can produce BHs up to ${\sim}45\,\Msun$ (assuming median values for the nuclear reaction rates, \citealt{2020ApJ...888...76M}). 

In summary, many sources of uncertainty affect our knowledge of the PI mass gap, the two principal ones being nuclear reaction rates ($\Delta m\approx 16\,\Msun$) and the collapse of the hydrogen envelope ($\Delta m\approx 20\,\Msun$). The combination of the main uncertainties has yet to be studied; therefore, it is not clear whether their effects sum up, as needed to explain the primary mass of \ThisEvent.  Based on these considerations, $\sim{}60-65$ M$_\odot$ is probably a conservative lower limit to the edge of the mass gap.

\subsection{Hierarchical Merger Scenario}
\label{sec:hierarchical}

In this section, we discuss general properties of hierarchical mergers, i.e.\ mergers involving one or two second-generation BHs;
we then present a Bayesian estimate of the relative probabilities (odds) that \ThisEvent is the product of a hierarchical BH merger, as opposed to a merger of first-generation (hereafter 1g) BHs. 
The masses of 2g BHs can fall in the PI gap: the merger remnant of GW170729 \citep{LIGOScientific:2018mvr} is a previously reported example. 
A 2g BH is alone at birth, unless it was a member of a triple system that remains bound after the formation of the BH. However, a single 2g BH might  acquire a companion through dynamical exchanges, if it forms in a dense cluster \citep{2002MNRAS.330..232C,2003ApJ...599.1260C,2019PhRvD.100d3027R}, or through migration-mediated interactions in an AGN disk \citep{2012MNRAS.425..460M,2017ApJ...835..165B}.

When formed, a 2g BH is subject to a
relativistic gravitational recoil velocity (kick) which can eject it from its birthplace \citep{2004ApJ...607L...9M}: see Fig.~\ref{fig:kick_mag}. For example, if a 2g BH forms in a dense star cluster, it might be retained in the cluster and subsequently acquire a new companion only if the local escape velocity is larger than ${\gtrsim}50$\,km\,s$^{-1}$ \citep{2008ApJ...686..829H,2016ApJ...824L..12O,2016ApJ...831..187A,2019PhRvD.100d1301G}. In this scenario, the companion might be a 1g BH or, less likely, a 2g BH \citep{2019PhRvD.100d3027R}.

The relativistic kick at birth strongly depends on BH spin: maximally spinning BHs with spins in the orbital plane and counter-aligned receive the largest kicks (e.g.,\ \citealt{2007ApJ...659L...5C,2007PhRvL..98w1102C}). Under the simplified assumption that all 1g BHs are born with spin $\chi=0$, then ${\sim}60$\% of 2g BHs are retained in a typical globular cluster, while if all 1g BH have $\chi=0.5$, then less than 3\% of merger products are retained \citep{2019PhRvD.100d3027R}. 
For these reasons, hierarchical mergers are expected to be more likely in nuclear star clusters (e.g.,\ \citealt{2016ApJ...831..187A,2020arXiv200307409A}), which have the highest escape velocity, rather than globular clusters or smaller stellar systems, such as open clusters and young star clusters.

Binaries with 2g BHs have several distinctive properties. The \emph{mass of the primary} can be significantly larger than the lower boundary of the PI mass gap, although the exact distribution strongly depends on the 1g mass function \citep{2019PhRvD.100d1301G,2019PhRvD.100d3027R,2019PhRvD.100j4015C,2020ApJ...893...35D}. 
Concerning the predicted \emph{mass ratios}, 1g+2g mergers tend to have more unequal mass ratios than 2g+2g mergers \citep{2019PhRvD.100d3027R}. The preference of \ThisEvent for a mass ratio $q\sim{}1$ suggests that this system is more likely consistent with a 2g+2g merger than with a 1g+2g merger.  \editNew{Indeed it is also conceivable for the system to be a merger of 1g BH where one component was \emph{above} the PI mass gap, though this is disfavored by the preference for near-equal masses (see Fig.~\ref{fig:NR}). }

\emph{Spin} measurements are a distinguishing feature of 2g BHs, because merger remnants are, on average, rapidly rotating with $\chi \sim 0.7$ (e.g.,\ \citealt{Baker:2003ds,2008PhRvD..77b6004B,Lousto:2009ka,2016ApJ...825L..19H,2017ApJ...840L..24F}). Since they acquire companions through dynamical exchanges, spins are expected to be isotropically distributed with respect to the BBH orbital angular momentum (e.g.,\ \citealt{2010CQGra..27k4007M,2016ApJ...832L...2R}). A direct consequence of spatial isotropy is that the predicted distribution of effective spins is 
broad and symmetric about $\chi_{\rm eff}=0$ \citep{2019PhRvD.100d1301G,2019PhRvD.100d3027R,2020RNAAS...4....2K,2018ApJ...854L...9F}. This prediction is broadly consistent with the range of $\chi_{\rm eff}$ and relatively large $\chi_{\rm p}$ values inferred for \ThisEvent, reported in Section~\ref{sec:PE_components}. 

Finally, to estimate the predicted \emph{merger rate} of PI BBHs from the hierarchical scenario, we should take into account that these mergers come from all kinds of star clusters with $v_\mathrm{esc}>50$\,km\,s$^{-1}$, hence both massive globular clusters and nuclear star clusters (or AGN disks). 
Considering only globular clusters and assuming that 1g BHs have zero spin, \cite{2019PhRvD.100d3027R} find that  ${\sim}3$\% of all mergers from globular clusters at redshift $z < 1$ have a component greater than $55\,\Msun$, corresponding to 7\% of the detectable BBHs from globular clusters.

\subsubsection{Bayesian analysis of hierarchical formation in globular clusters}
\label{subsec:hier_analysis}

We apply the method of \citet{2020arXiv200500023K} to recover posteriors over the relative rates of 1g+2g and 2g+2g mergers to 1g+1g mergers, as well as the odds ratios that the \ThisEvent binary is of 1g+2g or 2g+2g origin, using our estimates of source component masses and spins \editNew{within a physical model of the hierarchical merger process adapted to globular clusters.  This analysis illustrates the many physical parameters and associated uncertainties that enter any inference on the origin of massive BBH.}
\begin{figure}[tb!]
\begin{center}
\includegraphics[width=0.9\columnwidth]{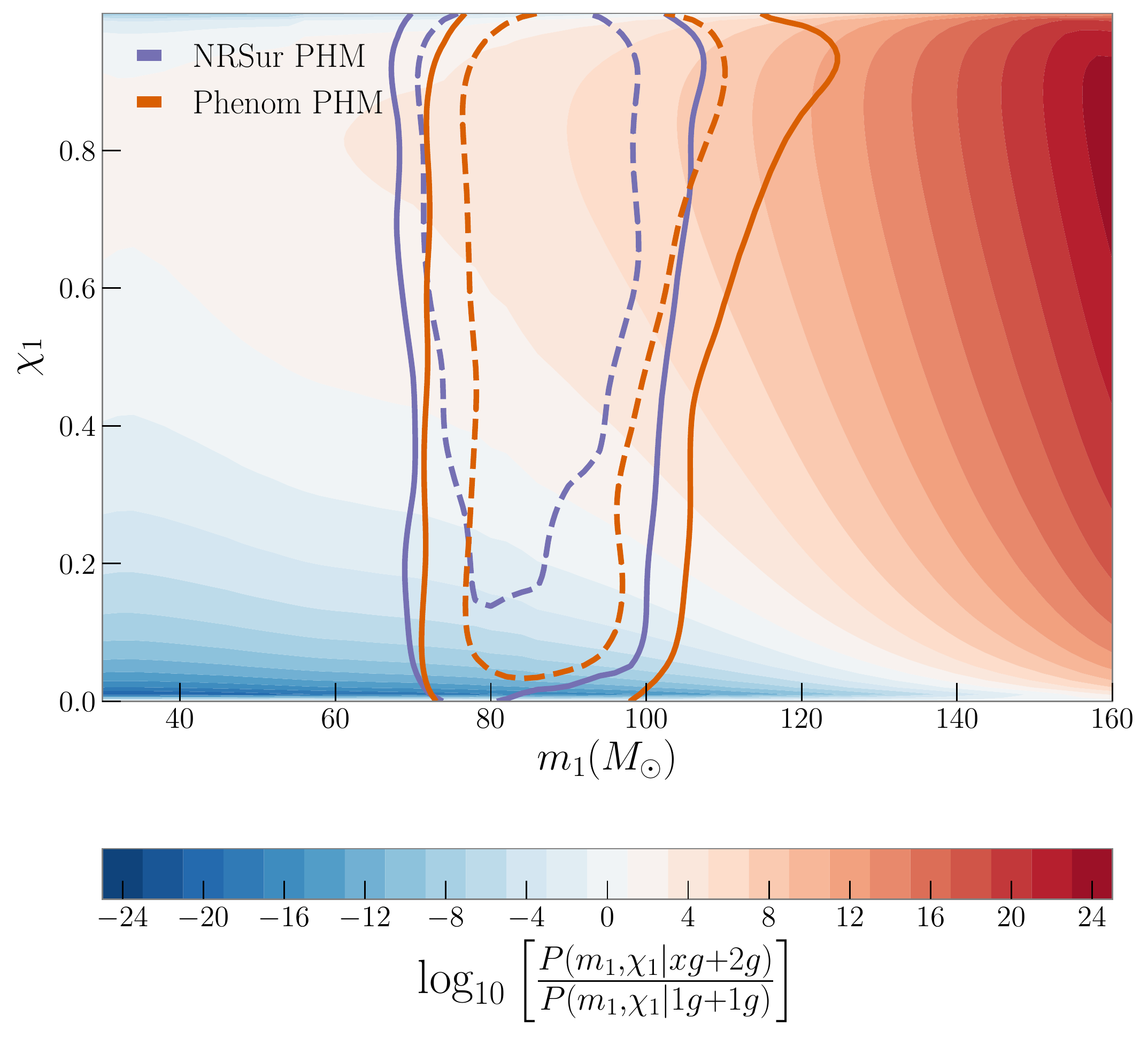}
    \caption{Marginalized Bayes factor for a BBH to be a second-generation merger (1g+2g or 2g+2g) as opposed to a first-generation BH merger, as a function of primary mass and spin, \editNew{in the globular cluster analysis of Section~\ref{subsec:hier_analysis} using a physically-motivated prior cutoff on 1g BH masses. The Bayes factor contours correspond to component masses and spins inferred using the NRSur~PHM model for \ThisEvent, and differ only slightly from those found using the Phenom~PHM model.} We show the 90\% and 68\% posterior credible regions for \ThisEvent as solid and dashed contours, respectively, \editNew{for both the NRSur~PHM and Phenom~PHM models}.} 
    \label{fig:hierarchical_bayes_factor}
    \end{center}
\end{figure}

We perform hierarchical Bayesian inference on a population model containing 1g+1g, 1g+2g and 2g+2g mergers. For the first-generation BBH population, we adopt Model C in \citet{2019ApJ...882L..24A}: the mass distribution is a mixture of a truncated power law and a high-mass Gaussian component \citep{MonashMass} while the component spins follow non-singular beta distributions with isotropic orientations. 
We also apply a theoretically motivated prior to the upper cutoff of the power-law mass component
$m_{\rm max}$ to reflect expectations from stellar collapse dynamics, in particular PI (see Section~\ref{sec:PI_uncertainty}).  Considering uncertainties due to nuclear reaction rates \citep{2019ApJ...887...53F} and fallback dynamics \citep{2020ApJ...888...76M} we take a prior Gaussian distribution over $m_{\rm max}$ with mean $50\,\Msun$ and standard deviation $10\,\Msun$. 
 \editNew{As mentioned above, we might also consider a 1g component \emph{above} the PI mass gap, i.e.\ of ${\sim}\,120$\,\Msun, however this would add complexity to the analysis while being unlikely to significantly alter the main outcome. }

For simplicity, all 1g+1g mergers are here assumed to occur in environments which could potentially lead to subsequent mergers \citep{2020ApJ...893...35D}. 
The 1g+2g and 2g+2g populations are formed by applying transfer functions from \citet{2020arXiv200500023K} to the 1g+1g distributions, motivated by simulations reported in \citet{2019PhRvD.100d3027R}.  The populations are combined into a mixture model by calculating branching ratios for the fraction of 1g+1g merger remnants that are retained in their cluster environments after receiving relativistic
kicks at merger (calculated with the \textsc{precession} package of \citealt{Precession,PrecessionCode}). 
The retention probability depends 
on the cluster escape velocity: as in \citet{2020arXiv200500023K}, we 
take a nominal cluster mass of $5\times 10^5\,\Msun$ with a Plummer radius of $1$\,pc \citep{1996AJ....112.1487H,2019arXiv191100018K}; to quantify dependence on the 
escape velocity, we also consider a higher mass cluster of $10^8\,\Msun$ representative of nuclear cluster environments.  

The overall retention fraction 
is also highly sensitive to the 1g+1g spin magnitude distribution. 
\citet{Fuller:2019sxi} find that a significant fraction of stellar BHs should form with near-zero natal spins; for BBH formed in clusters via dynamical interaction we may also neglect any possible effect of strong stellar binary interactions on 1g BH spins. 
Thus, we also perform an analysis with an expanded model that includes an additional sub-population of 1g BHs with zero spin, making up a fraction $\lambda_0$ of 1g+1g binary components.  We quote results both with and without this zero-spin channel. 
\begin{figure}[tb!]
    \begin{center}
    \includegraphics[width=0.99\columnwidth]{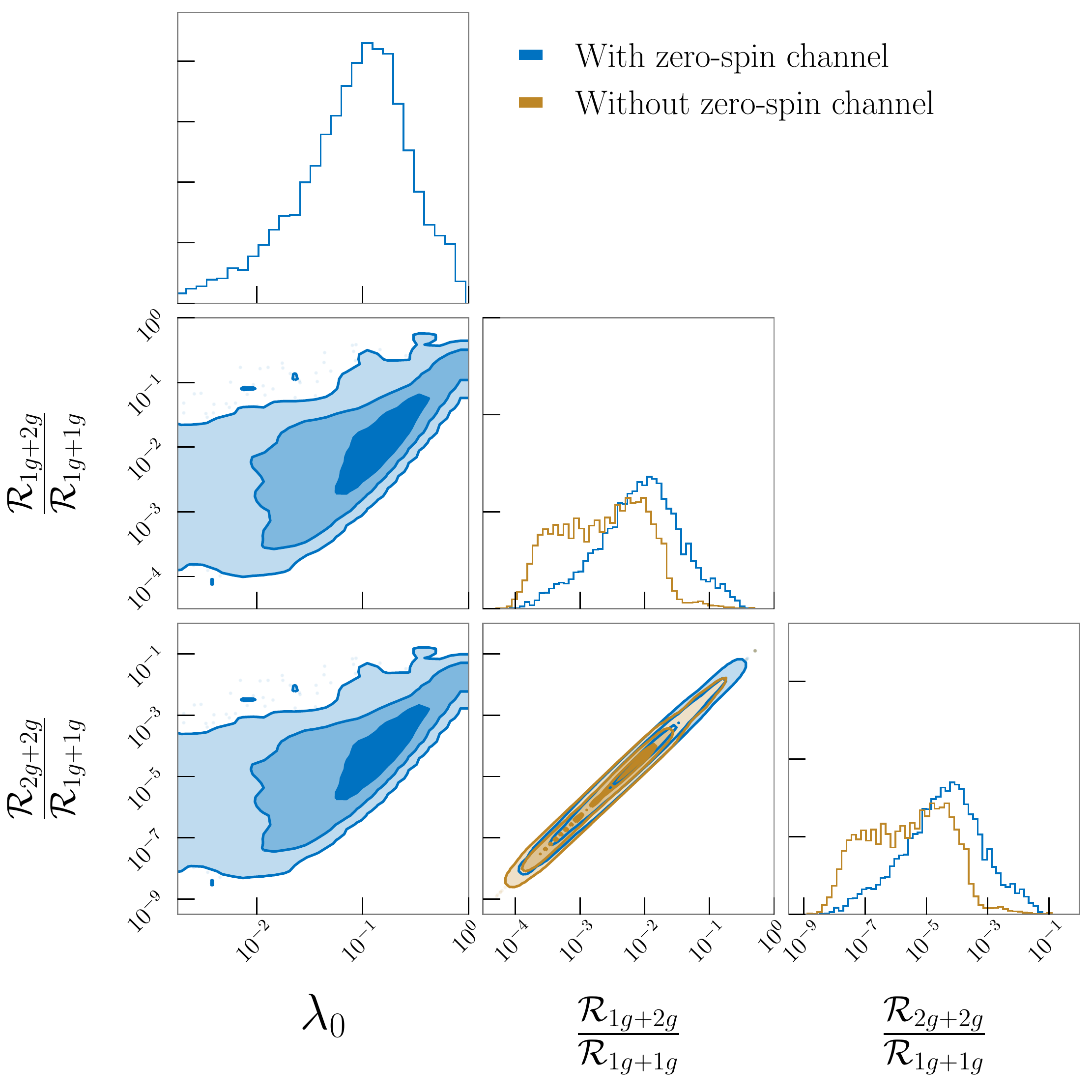}
    \caption{Relative rates of 1g+2g and 2g+2g as compared to first-generation mergers, \editNew{in globular cluster models} with (blue) and without (orange) a zero-spin stellar BH population (see Section~\ref{subsec:hier_analysis}), using \ThisEvent source parameters derived from NRSur~PHM.  In the model with zero-spin population, we also plot the fraction $\lambda_0$ of 1g+1g binary components belonging to this population.} 
    \label{fig:relative_rates_both}
    \end{center}
\end{figure}
\begin{table}[tb!]
\caption{Odds ratios for the source of \ThisEvent being a xg+2g merger vs.\ 1g+1g merger, \editNew{in the globular cluster models of Section~\ref{subsec:hier_analysis}}.}
\begin{ruledtabular}
    \begin{tabular}{l l l}
    No zero-spin channel & NRSur~PHM & Phenom~PHM \\
    \hline
        $P(\mathrm{1g\!+\!2g})$:$P(\mathrm{1g\!+\!1g})$ & \ORNRSurNoZeroHalf:1 & \ORIMRPHMNoZeroHalf:1 \\
    \rule{0pt}{3ex}%
        $P(\mathrm{2g\!+\!2g})$:$P(\mathrm{1g\!+\!1g})$ & \ORNRSurNoZeroTwo:1 & \ORIMRPHMNoZeroTwo:1 \\
    \hline
    With zero-spin channel & NRSur~PHM & Phenom~PHM \\
    \hline
        $P(\mathrm{1g\!+\!2g})$:$P(\mathrm{1g\!+\!1g})$ & \ORNRSurHalf:1& \ORIMRPHMHalf:1 \\
    \rule{0pt}{3ex}%
        $P(\mathrm{2g\!+\!2g})$:$P(\mathrm{1g\!+\!1g})$ & \ORNRSurTwo:1 & \ORIMRPHMTwo:1 \\
    \hline
    \end{tabular}
\end{ruledtabular}
    \label{tab:oddsratios}
\end{table}

We apply our mixture models to a set of BBH observations consisting of \ThisEvent and all BBH detections from O1 and O2, performing inference on model hyperparameters using \textsc{GWPopulation} \citep{2019PhRvD.100d3030T} and Bilby \citep{Ashton:2018jfp}. 
In Figure~\ref{fig:hierarchical_bayes_factor}, we plot the resulting marginalized Bayes factor of a given BBH being of xg+2g vs 1g+1g origin as a function of primary component mass and spin, together with the 90\% and 68\% credible regions for \ThisEvent.  Overall Bayes factors including dependence on both component masses and spins generally favor 2g+2g origin; the parameter-estimation likelihood strongly favors near-equal component masses, thus 1g+2g origin implying significantly unequal masses is disfavored by the data. 

Figure~\ref{fig:relative_rates_both} (orange contours) shows the relative rates of 1g+2g and 2g+2g mergers for the model without a 1g zero-spin sub-population. We find $\log_{10}\big(\mathcal{R}_{1g+2g}/\mathcal{R}_{1g+1g}\big)$ and $\log_{10}\big(\mathcal{R}_{2g+2g}/\mathcal{R}_{1g+1g}\big)$ to be $\RelRateNoZeroHalfMedian^{+\RelRateNoZeroHalfPlus}_{-\RelRateNoZeroHalfMinus}$ and $\RelRateNoZeroTwoMedian^{+\RelRateNoZeroTwoPlus}_{-\RelRateNoZeroTwoMinus}$, respectively.  
Applying the priors given by the relative rates and marginalizing over system parameters and population hyperparameters, we calculate the odds ratio that the source of \ThisEvent is a hierarchical merger, versus a 1g+1g merger.  Our results, summarized in the top half of Table~\ref{tab:oddsratios}, are 
dependent on the waveform chosen to produce the parameter estimation samples, as well as assumptions on the 1g+1g spin and mass ratio distributions.
Under these assumptions we find that 
\ThisEvent is most likely to be of 1g+1g origin, because the merger remnants of BBH with large component spins are often subject to kicks 
which eject them from low-mass clusters.  \editNew{Accounting for a possible 1g BH component \emph{above} the mass gap, implying a significantly unequal-mass merger, would be unlikely to alter this conclusion.} 

Applying the model with a 1g zero-spin sub-population yields higher retention fractions, due to lower merger recoil velocities, and higher relative hierarchical merger rates, shown in Fig.~\ref{fig:relative_rates_both} (blue contours): $\log_{10}\big(\mathcal{R}_{1g+2g}/\mathcal{R}_{1g+1g}\big)$ and $\log_{10}\big(\mathcal{R}_{2g+2g}/\mathcal{R}_{1g+1g}\big)$ become $\RelRateHalfMedian^{+\RelRateHalfPlus}_{-\RelRateHalfMinus}$ and $\RelRateTwoMedian^{+\RelRateTwoPlus}_{-\RelRateTwoMinus}$, respectively. When we include the zero-spin formation channel the odds ratios for hierarchical origin, summarized in the bottom half of Table~\ref{tab:oddsratios} increase by factors of ${\sim}2-10$, though we still favor \ThisEvent being due to a 1g+1g merger, both for the NRSur~PHM and Phenom~PHM parameter estimates.  In this case, the estimated fraction $\lambda_0$ of merging 1g BHs from the zero-spin channel is in the range ${\sim}\LambdaChiLow$--$\LambdaChiHigh$. 

We also consider a higher cluster mass of $10^8\,\Msun$ which may be representative of nuclear cluster enviroments: the higher cluster mass increases the odds of hierarchical origin by 3-4 orders of magnitude, for both models with and without a zero-spin formation channel.  For AGN environments essentially all 1g merger products are retained: however quantitative assessment
would require more detailed modelling of BH mergers in AGN beyond their increased escape velocity, and the 1g+2g and 2g+2g population models we present here are tuned to globular cluster models. 

In summary, the probability that \ThisEvent is due to a hierarchical merger in a stellar cluster is strongly dependent on the properties of 1g BHs in such environments, primarily on their \editNew{mass and} spin distributions and on the cluster escape velocity.  With significantly larger event samples it may be possible to disentangle the different model parameters: thus, similar, high mass BBH mergers constitute a future laboratory for BH populations and dynamics in cluster environments.

\subsection{Stellar merger scenario}
If a star grows an over-sized hydrogen envelope with respect to its helium core, it might directly collapse to a BH with mass ${\sim}60-100\,{}\Msun$ without entering the PI/PPI regime \citep{2019MNRAS.485..889S,2019MNRAS.487.2947D,2019ApJ...886...25B}. This scenario assumes that most of the hydrogen envelope collapses to a BH (see \citealt{2016ApJ...821...38S} for a discussion of the uncertainties). For a star to develop this over-sized hydrogen envelope, one or more mergers between a helium-core giant and a main-sequence companion are required \citep{2019arXiv191101434D}. As in the hierarchical-merger scenario, a BH born from a stellar merger is alone at birth, unless it was a member of a triple system that remains bound. In the field, the BH remains alone, while in a young star cluster, a globular cluster or a nuclear star cluster the BH can acquire a companion through exchanges.  \editNew{A further variation of this model is that \ThisEvent formed in a triple or multiple stellar system. The inner binary star of the triple might have merged producing the primary BH via the stellar merger mechanism described above. If the outer member is massive enough, it might then collapse becoming the secondary BH. The most critical aspect of a triple origin for \ThisEvent is the orbit of the tertiary body.} 

The \emph{primary mass} distribution predicted by the \editNew{stellar merger} model scales approximately as $m^{-5}$ with $m>60\,\Msun$. 
The expected \emph{mass ratio} is most likely $q\sim{} 0.4-0.6$, but all mass ratios between $q \sim 0.04$ and $q \sim 1$ are possible \citep{2019arXiv191101434D}. Hence, the primary mass and the mass ratio predicted by this scenario are compatible with the properties of \ThisEvent.

If the angular momentum of the core of the massive progenitor is not dissipated efficiently, this scenario predicts a high primary \emph{spin}. Since this is a dynamical formation scenario, spin orientations are expected to be isotropically distributed. As we already discussed for the hierarchical scenario (Section~\ref{sec:hierarchical}), an isotropic spin distribution is consistent with the fact that \ThisEvent has a low $\chi_{\rm eff}$ and a relatively large $\chi_{\rm p}$. 

Finally, \cite{2019MNRAS.487.2947D} predict that up to ${\sim}2\%$ of all BBH mergers from young star clusters involve BHs in the PI mass gap born from (multiple) stellar mergers. These represent ${\lesssim}10\%$ of all detectable mergers from star clusters. As noted in \cite{Jani_2020}, ${\lesssim}0.8$\% of all massive stars contribute to a BBH merger population in the PI mass-gap. 

In summary, the primary mass, mass ratio, effective spin and precession spin parameters of \ThisEvent are consistent with the stellar merger scenario in star clusters. The key difference between the hierarchical merger scenario and the stellar merger scenario is that the latter does not imply relativistic kicks at birth, hence it might be at least one order of magnitude more common in star clusters. On the other hand, the details of the stellar merger scenario depend on delicate assumptions about stellar mergers (e.g. that most mass is retained by the merger product) and massive star evolution (e.g. stellar rotation).

\subsection{AGN disk scenario}\label{sec:AGN}
The nucleus of an active galaxy might harbor tens of thousands of stellar-mass BHs that moved into the innermost parsec due to mass segregation \citep{1977ApJ...216..883B, 1993ApJ...408..496M,2000ApJ...545..847M,2014ApJ...794..106A,2018Natur.556...70H,2018MNRAS.478.4030G}. In this dense gaseous environment, BH orbits are efficiently torqued by gas drag till they align with the AGN disk \citep{2017ApJ...835..165B,2018ApJ...866...66M}. Once in the disk, BHs can accrete gas, and are expected to acquire companions and efficiently merge with them as an effect of gas torques \citep{2012MNRAS.425..460M,2014MNRAS.441..900M,2016ApJ...819L..17B,2017ApJ...835..165B,2019arXiv190704356M}. When these BHs merge with other BHs, there is a high chance that they are not ejected, because of the high escape velocity in galactic nuclei (thousands of km s$^{-1}$, given the proximity to a SMBH). Hence, AGN disks might easily harbor 2g BHs \citep{2019PhRvL.123r1101Y}.

As a consequence of the joint contribution of gas accretion, BBH mergers and high escape velocities, a fraction of BHs in AGN disks are expected to have masses in the PI mass gap or even in the IMBH regime \citep{2012MNRAS.425..460M}. Recent work suggests that they have a mass function similar to a power law, but significantly flatter than field BHs \citep{2019ApJ...876..122Y}, or even reminiscent of a broken power law \citep{2018ApJ...866...66M,2019PhRvL.123r1101Y}. The possible \emph{mass ratios} are highly uncertain, though most mergers have mass ratios less extreme than 10:1 \citep{2018ApJ...866...66M,2019PhRvL.123r1101Y}. 

As to the {\emph{spin}} magnitudes, we expect large spins ($\chi \sim 0.7$) if BHs in AGN disks are 2g BHs. 
Since BH orbits tend to align with the AGN disk, there might be some preferential alignment in the spins of BBHs \citep{2012MNRAS.425..460M}. 
Unlike the other possible scenarios, BBH mergers in an AGN disk may have an associated electromagnetic counterpart 
(an ultra-violet flare; \citealt{2019ApJ...884L..50M}). No \editNew{prompt} electromagnetic counterpart has been reported for \ThisEvent (e.g., \citealt{2019GCN.24619....1L,2019GCN.24624....1D,2019GCN.24623....1C,2019GCN.24620....1D,2019GCN.24625....1N,2019GCN.24644....1W,2019GCN.24657....1T,2019GCN.24658....1T,2019GCN.24646....1B}), \editNew{but see \cite{PhysRevLett.124.251102}, which appeared during the revision of this work, for a possible candidate optical counterpart.}

Finally, the fraction of BBHs in the PI mass gap born in AGN disks might be large: \cite{2019PhRvL.123r1101Y} suggest that 40\% of all AGN-assisted mergers detected by LIGO-Virgo will include a BH with mass $\gtrsim 50\,\Msun$; on the other hand, the overall contribution of AGN disks to the BBH merger rate might be low. Most works suggest a total merger rate ${\sim}1-10$\,Gpc$^{-3}$\,yr$^{-1}$ (e.g.,\  \citealt{2017ApJ...835..165B,2017MNRAS.464..946S,2019ApJ...876..122Y,2020arXiv200204037A}), while \cite{2018ApJ...866...66M} and \cite{2019arXiv191208218T} try to quantify additional uncertainties and end up with a merger rate ${\sim}10^{-3}-10^4$\,Gpc$^{-3}$\,yr$^{-1}$ and ${\sim}0.02-60$\,Gpc$^{-3}$\,yr$^{-1}$, respectively.

\section{Alternative Scenarios}

The short duration and bandwidth of the signal observed in the data open the possibility that the source may not be uniquely explained as an unusually massive, quasi-circular BBH merger formed via astrophysical processes. 
Furthermore, given a transient GW signal for which a full inspiral-merger-ringdown morphology is not directly evident in the data, 
a natural question is whether the signal may be consistent with other types of modelled GW source. 
We consider several possible alternative scenarios below; all are disfavored either by the data, or by low prior probability of the alternative hypothesis, or by both.

\subsection{Eccentricity and Head-On Collisions}
\label{sec:alt_ecc} 
The waveform observed in the data may also be consistent with the merger of a BBH with non-zero orbital eccentricity. The short duration of the signal makes it difficult to distinguish the amplitude modulation associated with precession from that due to eccentric orbits, or even head-on collisions \editNew{(mergers that happen immediately at closest approach)} \citep{CalderonBustilloHeadOn}. More quantitative evaluation of this issue requires further development of accurate and computationally efficient eccentric waveforms, and is thus deferred to future work. Efforts to efficiently detect and estimate the parameters for eccentric compact binaries are under development \citep{PhysRevD.98.083028,10.1093/mnras/stz2996,lenon2020measuring}. 

From an astrophysical perspective, BBHs with eccentricity $e\gtrsim{}0.1$ at merger are deemed to be 
almost impossible from isolated binary evolution, but might account for ${\sim}1$\% of all BBH mergers in globular clusters and other dense star clusters 
(\citealt{2006ApJ...640..156G,2014ApJ...784...71S,2018ApJ...855..124S,2017ApJ...840L..14S,2018PhRvD..97j3014S,2018PhRvD..98l3005R,2018MNRAS.481.5445S,2019ApJ...871...91Z,2020arXiv200211278F}). 
 \editNew{Head-on collisions are even more rare due to the small geometric cross-section of individual BHs \citep{2014ApJ...784...71S}.}

The high concentration of BHs in nuclear star clusters might produce a population of extremely eccentric BBHs from single--single BH capture by GW radiation \citep{2009MNRAS.395.2127O,2018ApJ...860....5G}, although the cross-section for this process is orders of magnitude lower than the cross-section for a binary -- single encounter \citep{2014ApJ...784...71S}. Finally, Kozai-Lidov resonances \citep{1962AJ.....67..591K,1962P&SS....9..719L} in triple systems might also trigger eccentric BBHs  in the field \citep{2014MNRAS.439.1079A,2017ApJ...836...39S,2017ApJ...841...77A} or in dense star clusters \citep{2003ApJ...598..419W,2012ApJ...757...27A,2016ApJ...816...65A,2016MNRAS.463.2443K,2018ApJ...856..140H}. Considering that ${\sim}25$\% of massive stars are members of triple systems \citep{2014ApJS..215...15S}, \cite{2017ApJ...841...77A} estimate a merger rate of ${\sim}0.3-2.5$\,Gpc$^{-3}$\,yr$^{-1}$ from isolated resonating triple systems, up to ${\sim}5$\% of which retain high eccentricity at merger.

Most of the formation channels that can explain the mass of the primary BH in \ThisEvent  are based on dynamical encounters in dense star clusters  and allow for the formation of eccentric BBHs. Hence, even if eccentric BBH mergers are estimated to be extremely rare, we cannot exclude that the source binary of \ThisEvent has non-zero eccentricity.

\subsection{Strong Gravitational Lensing}

Another possible explanation for the apparent inconsistency of \ThisEvent  with formation from stellar collapse is gravitational lensing of the signal by galaxies or galaxy clusters \citep{Ng:2017yiu,2018arXiv180205273B,2018MNRAS.480.3842O,Li:2018prc,2018MNRAS.475.3823S}. 
If \ThisEvent is a strongly lensed signal, it will receive a magnification $\mu$ 
defined such that GW amplitude is increased by a factor $\mu^{1/2}$ relative to the unlensed case. 
For a given observed GW signal, the luminosity distance to the source, and thus its redshift, may then be much larger than in the absence of lensing. 
Strong lensing is expected to be relatively rare at current detector sensitivities, with one in every hundred to thousand detected events 
strongly lensed by individual galaxies \citep{1998PhRvD..58f3501H,Ng:2017yiu,2018MNRAS.480.3842O,Li:2018prc} and a similarly low rate for lensing by galaxy clusters \citep{2018MNRAS.475.3823S,2019RPPh...82l6901O}.

Under the strong lensing hypothesis $\mathcal{H}_L$, we assume the event comes from a BBH population of stellar collapse origin, and infer its magnification, redshift and component masses.  The posterior distribution of the parameters is \citep{Pang:2020qow}
\begin{equation}
    p(\mu, \vartheta|d,\mathcal{H}_L) \propto p(d|\vartheta)\, p(\vartheta|\mu,\mathcal{H}_L)\, p(\mu|\mathcal{H}_L),
\end{equation}
where $\vartheta$ are the apparent parameters of the waveform received at the detector, which differ from the source-frame parameters due to effects of redshift and lensing. 
For the prior over component masses and redshift of the source $p(m_1,m_2,z|\mathcal{H}_L)$, we take a binary BH population model from those used in \citet{2019ApJ...882L..24A} for O1 and O2 observations, with fixed population parameters $\lambda=0$, $\alpha=1$, $\beta_q=0$, $m_{\rm min}=5\,\Msun$, multiplied by the optical depth of lensing by galaxies $\tau(z)=4.17\times 10^{-6} (D_c(z)/{\rm Gpc})^3$ \citep{Haris:2018vmn}, where $D_c$ is the comoving distance.  For the primary mass upper cutoff $m_{\rm max}$ we consider two different values to account for uncertainties in the edge of the PISN gap, $m_{\rm max} = (50,65)\,\Msun$. 
We use the lensing prior $p(\mu|\mathcal{H}_L)\propto \mu^{-3}$ \citep{1986ApJ...310..568B} with a lower limit $\mu>2$ appropriate to strong lensing. We adopt the NRSur~PHM waveform model. The resulting magnification estimate is mostly sensitive to the measurement of the component masses. 

We find the required magnification under the lensing hypothesis, taking $m_{\rm max} = 50\,\Msun$ ($m_{\rm max} = 65\,\Msun$), is $\lensingmagnification$ ($\lensingmagnificationb$), with source-frame masses $\lensingMp\,{}\Msun$ ($\lensingMpb\,{}\Msun{}$), $\lensingMs\,{}\Msun$ ($\lensingMsb\,\Msun$) at redshift $\lensingz$ ($\lensingzb$). 
At these redshifts, the lensing optical depth is low: $\lensingtau$ ($\lensingtaub$). 

One possible signature of strong lensing would be multiple GW images, which may give rise to two events occurring closer in time than expected for a Poisson process of BBH mergers, and with consistent sky localization and intrinsic parameters.  Another candidate GW signal, \OtherEvent \citep{GCN24632}, was reported $4.6$ hours after \ThisEvent, however 
the two events' sky localizations strongly disfavor lensing, showing no overlap \citep{Haris:2018vmn,Hannuksela:2019kle,Singer:2019vjs}. 
Any lensed counterpart image could, though, have been too weak to be confidently detected, and may have arrived at a larger time-separation or when the detectors were not operating. Moreover in the case of galaxy cluster lensing, the counterpart could appear at a separation of years in time \citep{2018MNRAS.475.3823S}. 

Given the low expected lensing rate and optical depth and the absence of an identifiable multi-image counterpart close to \ThisEvent, our current analyses find no evidence in favor of the strong lensing hypothesis.  Future analyses using sub-threshold searches and multi-image searches on all event pairs may yield better constraints on strong lensing \citep{Haris:2018vmn,Hannuksela:2019kle,2019arXiv190406020L,McIsaac:2019use}.

\subsection{Primordial BH Mergers}

Primordial BHs \citep[PBHs;][]{1974MNRAS.168..399C,Khlopov:2008qy} are thought to be formed from collapse of dark matter overdensities in the very early Universe (at redshifts $z > 20$, i.e. before the formation of the first stars), and may account for a nontrivial fraction of the density of the Universe \citep{2016PhRvD..94h3504C,Clesse:2016vqa}.  Since the binary components of \ThisEvent are unlikely both to have formed directly from stellar collapse, it is possible that they may be of PBH origin \citep{2016PhRvL.116t1301B}; however, theoretical expectations of the mass distribution and merger rate of PBH binaries have large uncertainties (e.g., \citealt{Byrnes:2018clq}), \editNew{so} we do not attempt to quantify such scenarios.  Some theories of PBH formation predict predominantly small component spins $\chi \ll 1$ \citep{Chiba:2017rvs}, which are disfavored by our parameter estimates for \ThisEvent (Section~\ref{sec:PE_components} and Fig.~\ref{fig:spins}).

\subsection{Cosmic String Signal Models}

Cosmic string cusps and kinks (e.g., \citealt{Damour:2000wa}) may yield short duration transient signals (bursts) with support at low frequency \citep[e.g.,][Section~IV~D]{MemoryEffectOrCosmicString}.  
There has been no previous detection of a cosmic string GW burst \citep{Abbott:2017mem,Aasi:2013vna} \editNew{and bounds on cosmic string model parameters derived from the overall contribution to the GW stochastic background are generally more stringent by orders of magnitude than bounds from direct burst searches \citep{Abbott:2017mem,LIGOScientific:2019vic}.  Thus detection of a burst signal in current data is a priori unlikely, however we consider this possibility for completeness.}
Here we estimate the likelihood for the data to be produced by a cosmic string signal and compare this with binary BH merger models as described in previous sections. 

\begin{figure}[tb!]
\begin{center}
\includegraphics[width=\columnwidth]{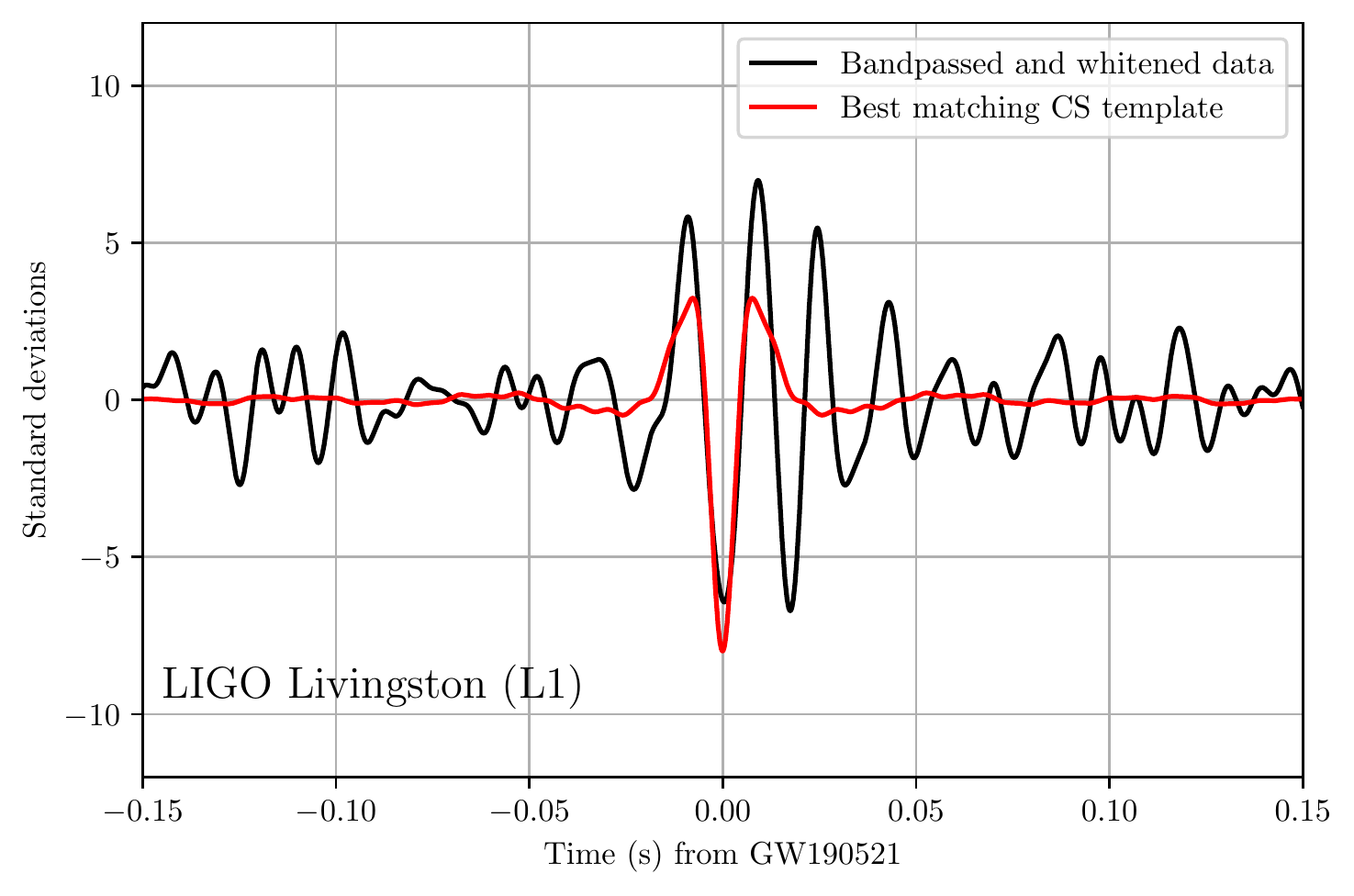}
\caption{The LIGO Livingston strain data time-series is shown in dark grey. The best-matching cosmic string cusp template is shown in red.  Both are whitened using the detector's noise spectrum; the strain data are additionally band-passed between 20\,Hz and 128\,Hz to suppress noise well above the frequency support of the template.}
\label{fig:cosmicstring}
\end{center}
\end{figure}

\ThisEvent is identified by the cosmic string matched filter search pipeline \citep{Abbott:2017mem,Aasi:2013vna}; however, the maximum signal-to-noise ratios in this search (${\sim}6$ and ${\sim}8$ in LIGO Hanford and Livingston, respectively) are much lower than for either modelled BBH search templates, or best-fit binary merger waveform models, or for unmodelled reconstructions, suggesting the data strongly prefers a binary merger model to a cosmic string or cusp.  In addition the signal appears inconsistent with the cosmic string template, as evidenced by large values of the $\chi^2$ test statistic, over $7$ ($3$) in LIGO Livingston (Hanford), while the expected value for a cosmic string signal is unity. Figure~\ref{fig:cosmicstring} illustrates the mismatch between the best matching cusp waveform and the Livingston data. 

As a refinement of this analysis accounting for parameter uncertainties and phase space volumes for different models, we 
compare the Bayes factors for a BBH waveform model to cosmic string cusp and kink models, respectively.  Since the true rates of binary mergers in the mass range of \ThisEvent and of cosmic string transient signals are both unknown, we do not attempt to quantify prior odds for the two hypotheses.  We use methods and models, including priors on cosmic string signal parameters, detailed in \citet{MemoryEffectOrCosmicString}.  For the BBH model, we use the NRSur~PHM waveform model with priors on component masses, spins and distance matching those in Section~\ref{sec:PE}. 
Using the \texttt{dynesty} nested sampling algorithm \citep{10.1093/mnras/staa278} within Bilby \citep{Ashton:2018jfp,romeroshaw2020bayesian}, we find Bayes factors $\log_{10} B_{\rm cusp}^{\rm BBH} = \logTenBFbbhVScusp$ and $\log_{10} B_{\rm kink}^{\rm BBH} = \logTenBFbbhVSkink$, strongly favoring the BBH hypothesis, thus it is highly unlikely that this event was of cosmic string origin.

\subsection{Core Collapse Supernova Signal Models}
Core-collapse supernova (CCSN) signals modeled in multidimensional simulations 
(e.g.,\ \citealt{Andresen2019,OConnorCouch2018,Powell2019,Yakunin2017}) are of long 
duration 
(up to 1.5\,s), may be broad-band (from few Hz to around 1-2\,kHz), and do not have morphology 
similar to \ThisEvent. Such signals are expected to be detectable for sources within a few 
tens of kpc distance from Earth \citep{Abbott2020}. We expect a CCSN signal to have 
electromagnetic and neutrino counterparts; however no counterpart has been observed within a few hundred kpc \citep{GCNallGW190521}.
GWs from extreme emission CCSNe, produced by highly deformed, rapidly rotating or fragmented 
cores, could be observed at a few tens of Mpc range \citep{Abbott2020}. At that distance we do not 
expect to detect neutrinos, however a detection of GWs produced by these extreme emission scenarios 
is overall very unlikely and an electromagnetic counterpart would still be expected. 
It is thus implausible that the \ThisEvent signal was produced by CCSN.

\section{Summary and Conclusions}

The GW signal \ThisEvent observed by the Advanced LIGO and Advanced Virgo detectors is consistent with emission from the coalescence of a high mass BBH system. Under that assumption, at least one of the component BHs in the binary has a mass in the PI mass gap with high probability, and the final merged BH has a mass in the IMBH range. There is no conclusive previous evidence from electromagnetic observations for the existence of IMBHs in this mass range (${\sim}100-1000\,\Msun$).

The signal is well described by waveforms derived from general relativity incorporating spin-induced orbital precession and higher-order multipoles, but neglecting orbital eccentricity. 
There is only weak evidence for the effects of orbital precession in the signal, and there is moderate but not conclusive evidence for high spin components in the orbital plane. 
 \editNew{The Bayes factor for the presence of higher-order multipoles is slightly negative, disfavoring their presence;} however 
their inclusion enables more precise estimates of the source's component masses, distance and inclination to the line of sight. 

Several tests demonstrate the consistency of the signal with GR predictions,  
including agreement between full inspiral-merger-ringdown waveforms derived from numerical relativity, and ringdown-only waveforms formed by summing the quasi-normal modes of the final Kerr BH. 
Because of the short duration and low bandwidth of the observed signal, possible sources other than a quasi-circular, high mass BBH merger may be considered.  Initial studies using waveforms that include orbital eccentricity suggest a partial degeneracy between eccentricity significantly greater than 0.1 and precession; however, significant eccentricity is disfavored by a low prior probability. 
The possibility that the signal is from a cosmic string cusp or kink is considered, but the signal is a poor match to waveform models from such sources. 
Finally, the possibility of the source being a core collapse supernova is considered; this is disfavored by the lack of any counterpart electromagnetic or neutrino signal for CCSN. 

We discussed various scenarios for the formation and evolution of such a massive system. Uncertainties on the PI mass gap might justify the formation of BHs with mass $>65\,\Msun$ 
from stellar collapse. Alternatively, the primary BH might be the result of the merger of two smaller BHs (hierarchical scenario), 
or of two massive stars. 
The mild evidence for precession suggests a dynamical formation scenario, which predicts nearly isotropic spin orientation, such as the hierarchical BH merger or the outcome of a stellar merger in star clusters. The formation of \ThisEvent via isolated binary evolution appears disfavored.

Finally, we considered the possibilities that \ThisEvent is a strongly lensed signal from a lower-mass, high redshift BBH merger, or that the system is a binary of primordial BHs (formed in the early Universe). 
The mass of the system, the mass ratio and the merger rate derived for this event are consistent with all of these scenarios, although the prior probability of strong lensing is low. 

It is not possible to conclude at this time whether \ThisEvent represents the first of a new population of BBHs, or is merely at the high-mass end of the population of BBH systems already observed by LIGO and Virgo. 
A future publication will address this question in the context of the larger sample of BBH events observed in the first half of the LIGO--Virgo O3 run, O3a (\editNew{2019 April 1 through October 1}). 
The answer, 
whether positive or negative, has the potential to provide new insights into the population and evolution of the most massive stars. We look forward to observing more binary mergers with high mass to further inform our understanding of these phenomena.

In upcoming future observing runs \citep{ObservingScenariosArxiv} we expect the global advanced detector network's sensitivity to BBH mergers to increase significantly, with potentially several hundreds of detections per year, reaching out to redshifts of a few or more.  We may thereby observe a large sample of events similar to \ThisEvent, which will constitute a unique source of information on the binary formation environments and channels, and on the dynamics and evolution of massive stellar BHs over cosmic history.  This will enable us to address questions such as the natal spin and mass distribution of BHs as a product of stellar collapse dynamics including PI; stellar cluster masses and dynamics; merger kicks due to GW emission; and tests of general relativity for highly spinning BHs. 

The evidence for high mass BBH and IMBH binaries advances the science case for enhanced sensitivity at lower GW frequencies ($\lesssim 5$~Hz). The proposed next-generation ground-based GW detectors, such as Einstein Telescope \citep{Punturo:2010zz} and Cosmic Explorer \citep{Evans:2016mbw,2019arXiv190704833R}, can observe events similar to \ThisEvent up to redshift ${\sim}20$ \citep{Gair:2010dx,Jani2020}. Such potential observations at high redshift can probe the dynamics and evolution of massive stellar BHs over cosmic history. \ThisEvent also motivates the possibility for multiband GW observations \citep{Fregeau:2006yz,Miller:2008fi,2016PhRvL.116w1102S, Jani2020} between ground-based detectors and upcoming space-based detectors like LISA \citep{2017arXiv170200786A}. Such joint observations will provide a unique opportunity to test general relativity with highly spinning BHs \citep{2016PhRvL.117e1102V} and probe binary formation environments and channels through measurements of spin evolution and eccentricity \citep{Cutler2019}.

\clearpage

\section{Acknowledgments}
The authors gratefully acknowledge the support of the United States
National Science Foundation (NSF) for the construction and operation of the
LIGO Laboratory and Advanced LIGO as well as the Science and Technology Facilities Council (STFC) of the
United Kingdom, the Max-Planck-Society (MPS), and the State of
Niedersachsen/Germany for support of the construction of Advanced LIGO 
and construction and operation of the GEO600 detector. 
Additional support for Advanced LIGO was provided by the Australian Research Council.
The authors gratefully acknowledge the Italian Istituto Nazionale di Fisica Nucleare (INFN),  
the French Centre National de la Recherche Scientifique (CNRS) and
the Netherlands Organization for Scientific Research, 
for the construction and operation of the Virgo detector
and the creation and support  of the EGO consortium. 
The authors also gratefully acknowledge research support from these agencies as well as by 
the Council of Scientific and Industrial Research of India, 
the Department of Science and Technology, India,
the Science \& Engineering Research Board (SERB), India,
the Ministry of Human Resource Development, India,
the Spanish Agencia Estatal de Investigaci\'on,
the Vicepresid\`encia i Conselleria d'Innovaci\'o, Recerca i Turisme and the Conselleria d'Educaci\'o i Universitat del Govern de les Illes Balears,
the Conselleria d'Innovaci\'o, Universitats, Ci\`encia i Societat Digital de la Generalitat Valenciana and
the CERCA Programme Generalitat de Catalunya, Spain,
the National Science Centre of Poland,
the Swiss National Science Foundation (SNSF),
the Russian Foundation for Basic Research, 
the Russian Science Foundation,
the European Commission,
the European Regional Development Funds (ERDF),
the Royal Society, 
the Scottish Funding Council, 
the Scottish Universities Physics Alliance, 
the Hungarian Scientific Research Fund (OTKA),
the French Lyon Institute of Origins (LIO),
the Belgian Fonds de la Recherche Scientifique (FRS-FNRS), 
Actions de Recherche Concertées (ARC) and
Fonds Wetenschappelijk Onderzoek – Vlaanderen (FWO), Belgium,
the Paris \^{I}le-de-France Region, 
the National Research, Development and Innovation Office Hungary (NKFIH), 
the National Research Foundation of Korea,
Industry Canada and the Province of Ontario through the Ministry of Economic Development and Innovation, 
the Natural Science and Engineering Research Council Canada,
the Canadian Institute for Advanced Research,
the Brazilian Ministry of Science, Technology, Innovations, and Communications,
the International Center for Theoretical Physics South American Institute for Fundamental Research (ICTP-SAIFR), 
the Research Grants Council of Hong Kong,
the National Natural Science Foundation of China (NSFC),
the Leverhulme Trust, 
the Research Corporation, 
the Ministry of Science and Technology (MOST), Taiwan
and
the Kavli Foundation.
The authors gratefully acknowledge the support of the NSF, STFC, INFN and CNRS for provision of computational resources.


{\it We would like to thank all of the essential workers who put their health at risk during the COVID-19 pandemic, without whom we would not have been able to complete this work.}

Strain data from the LIGO and Virgo detectors associated with \ThisEvent can be found in \citet{GW190521Data}.
This document has been assigned the LIGO document number LIGO-P2000021.

\clearpage

\bibliography{gw190521-implications-main}

\allauthors

\listofchanges
\end{document}